\documentclass[article, prl, nobibnotes, secnumarabic, superscriptaddress, twocolumn, aps,longbibliography]{revtex4-2}

\usepackage{amsmath,amssymb,graphicx}
\usepackage{upgreek}
\usepackage{bm}
\usepackage{xr-hyper}
\usepackage{hyperref}
\usepackage{xcolor}
\usepackage{graphicx}  
\hypersetup{colorlinks,breaklinks,
            urlcolor=[rgb]{0,0,0.64},
            linkcolor=[rgb]{0,0,0.64},
            citecolor=[rgb]{0,0,0.64},
            filecolor=[rgb]{0,0,0.64}}

\newcommand{\TCDW}{$T_{\mathrm{CDW}}$}

\makeatletter
\newcommand*{\addFileDependency}[1]{
  \typeout{(#1)}
  \@addtofilelist{#1}
  \IfFileExists{#1}{}{\typeout{No file #1.}}
}
\makeatother
\newcommand*{\myexternaldocument}[1]{
    \externaldocument{#1}
    \addFileDependency{#1.tex}
    \addFileDependency{#1.aux}
}
\myexternaldocument{supp_arXiv}

\begin{document}

\title{Revisiting the charge-density-wave superlattice of 1\!\textit{T}-TiSe$_\text{2}$}

\author{Wei Wang}
\thanks{These authors contributed equally to this work: W.W. and P.L.}
\affiliation{Condensed Matter Physics and Materials Science Division, Brookhaven National Laboratory, Upton, NY 11973, USA}
\affiliation{Department of Physics, University of Science and Technology of China, Hefei, Anhui 230026, People’s Republic of China}

\author{Patrick Liu}
\thanks{These authors contributed equally to this work: W.W. and P.L.}
\affiliation{Department of Applied Physics, Stanford University, Stanford, California 94305, USA}

\author{Lijun Wu}
\thanks{Correspondence to: ljwu@bnl.gov, alfredz@stanford.edu, and zhu@bnl.gov.}
\affiliation{Condensed Matter Physics and Materials Science Division, Brookhaven National Laboratory, Upton, NY 11973, USA}

\author{Jing Tao}
\affiliation{Condensed Matter Physics and Materials Science Division, Brookhaven National Laboratory, Upton, NY 11973, USA}
\affiliation{Department of Physics, University of Science and Technology of China, Hefei, Anhui 230026, People’s Republic of China}

\author{Genda Gu}
\affiliation{Condensed Matter Physics and Materials Science Division, Brookhaven National Laboratory, Upton, NY 11973, USA}

\author{Alfred Zong}
\thanks{Correspondence to: ljwu@bnl.gov, alfredz@stanford.edu, and zhu@bnl.gov.}
\affiliation{Department of Applied Physics, Stanford University, Stanford, California 94305, USA}
\affiliation{Department of Physics, Stanford University, Stanford, California 94305, USA}
\affiliation{Stanford Institute for Materials and Energy Sciences, SLAC National Accelerator Laboratory, Menlo Park, California 94025, USA}

\author{Yimei Zhu}
\thanks{Correspondence to: ljwu@bnl.gov, alfredz@stanford.edu, and zhu@bnl.gov.}
\affiliation{Condensed Matter Physics and Materials Science Division, Brookhaven National Laboratory, Upton, NY 11973, USA}
\affiliation{Department of Physics and Astronomy, Stony Brook University, Stony Brook, NY 11794, USA}

\date{\today}

\begin{abstract}
A number of intriguing phenomena, including exciton condensation, orbital ordering, and emergence of chirality, have been proposed to accompany charge-density-wave (CDW) formation in the layered transition metal dichalcogenide 1$T$-TiSe$_2$. Explaining these effects relies on knowledge of the atomic displacement pattern underlying the CDW, yet structural proposals based on spatially-averaging bulk crystal diffraction and surface-dependent scanning tunneling microscopy have remained inconsistent. Here, we revisit the CDW superlattice structure with selected-area electron diffraction, a bulk-sensitive probe capable of capturing sub-micrometer spatial variations while maintaining high momentum resolution. We resolved two distinct, spatially separated CDW phases characterized by different interlayer ordering. In both phases, previously reported atomic displacement patterns fail to account for the observed extinction rules. Instead, our analysis reveals a new superlattice structure, which features a large number of nearly degenerate CDW domains. These findings not only provide a new basis for understanding the gyrotropic electronic order and metastability in 1$T$-TiSe$_2$, they also underscore the importance of bulk-sensitive mesoscopic techniques in investigating materials that host unconventional superlattices.
\end{abstract}

\maketitle

A charge-density-wave (CDW) transition is characterized by a spontaneously broken crystalline translation symmetry, leading to a spatial modulation of the electron density and a periodically distorted nuclear lattice, known as the CDW superlattice. Many important classes of quantum materials, such as unconventional superconductors \cite{Frano2020,Rossi2022,Tam2022} and correlated insulators \cite{Hellmann2012,Liu2021}, feature one or multiple CDWs in their phase diagram. Hence, determining the precise atomic structure of the CDW superlattice serves as the foundational step for understanding their properties. For an incommensurate CDW, a unit cell is undefined, and one has to resort to a superspace group to describe the spatial symmetries. In that case, the periodic lattice distortion can only be approximated by a sinusoidal wave and its harmonics \cite{Malliakas2006}. For a commensurate CDW, there is a well-defined unit cell---known as the supercell---and a definite space group assignment. In principle, structural refinement can yield complete information of atomic positions in the supercell, from which the periodic lattice distortions can be deduced.

\begin{figure}[b!]
    \centering
    \includegraphics[width=1\columnwidth]{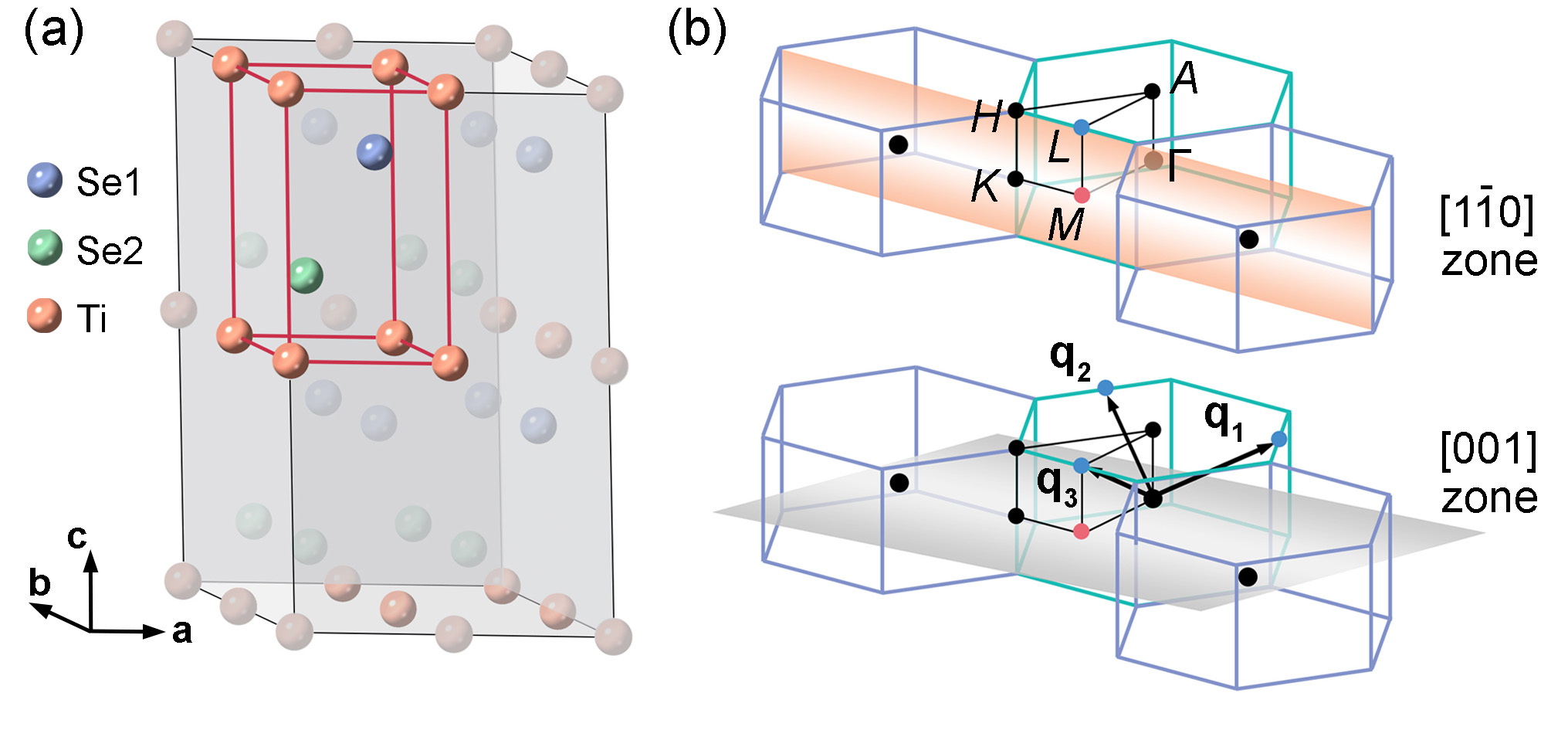}
    \caption{\textbf{Real space and momentum space unit cells.} (a)~Schematic crystal structure of 1$T$-TiSe$_2$ above \TCDW~(red prism with three atoms per unit cell: Se1, Se2, Ti) and below \TCDW~(the entire prism, showing a commonly adopted $2\times2\times2$ supercell with 24 atoms). (b)~The high-temperature Brillouin zone with high-symmetry points labeled, showing the two diffraction planes of the indicated zone axes that are used in the experiments. $\mathbf{q}_{1,2,3}$ are the three-fold symmetric CDW wave vectors at the $L$ points, as proposed in ref.~\cite{DiSalvo1976}.}
    \label{fig:Real_space_and_momentum_space_unit_cells}
\end{figure}

A prototypical commensurate CDW is found in a layered transition metal dichalcogenide, 1$T$-TiSe$_2$. Above $T_\text{CDW}\approx200$~K, its unit cell consists of one Ti and two Se atoms [Fig.~\ref{fig:Real_space_and_momentum_space_unit_cells}(a)], which is believed to transform into a $2\times2\times2$ supercell when cooled below the transition temperature. This commensurate CDW has been subjected to intense investigations due to its proximity to a superconducting dome \cite{Morosan2006,Kusmartseva2009}, the concomitant ordering in electron orbitals \cite{VanWezel2011,Peng2022}, the controversial role of exciton condensation in driving the CDW state \cite{Hughes1977,Suzuki1985,Kidd2002,Cercellier2007,Calandra2011,Porer2014,Kogar2017b,Burian2021,Lin2022,Cheng2022,Kurtz2024}, and the intriguing domain wall \cite{Duan2021,Cheng2024} and electronic properties \cite{Duan2023,Huber2024} induced by ultrashort laser pulses. In particular, it was proposed that the commensurate CDW is chiral, yet the origin of the chirality, if present, remains unsettled \cite{Ishioka2010, VanWezel2011,Castellan2013,Zenker2013,Gradhand2015,Lin2019,Xu2020,Jog2023,Nie2023,Kim2024,Xiao2024,Ueda2024,Qiu2025}. An accurate grasp of the superlattice structure is hence critical to explaining both the equilibrium properties and the non-equilibrium response.

Structural refinements of the CDW state, however, have been inconsistent among different X-ray and neutron scattering measurements. Although the high-temperature space group of 1$T$-TiSe$_2$ is generally assigned to $P\overline{3}m1$, the low-temperature space group has several proposals based on refinement studies: $P\overline{3}c1$ \cite{DiSalvo1976,Kitou2019}, $P\overline{3}m1$ \cite{Wegner2020}, and $P321$ \cite{Kim2024}. These discrepancies have important implications. For instance, $P321$ indicates a chiral phase \cite{Kim2024} while $P\overline{3}c1$ does not because the latter structure is centrosymmetric. There is no clear understanding of why the discrepancies arise, but given the relatively large beam spot in most X-ray and neutron refinement studies, one hypothesis is that the spatial heterogeneity of the CDW distortion in an otherwise nearly perfect parent lattice has been overlooked. 

In this Letter, we reexamine the CDW structure of 1$T$-TiSe$_2$ using selected-area electron diffraction in a transmission electron microscope, where a mesoscopic probing region (240~nm in diameter) allowed us to isolate domains of two distinct atomic arrangements in the sample bulk while maintaining high momentum resolution in the diffraction patterns. We name the two arrangements as $\mathcal{L}$ and $\mathcal{M}$ phases of the CDW, which feature $2\times2\times2$ and $2\times2\times1$ supercells, respectively. Through analysis of superlattice extinction rules and diffraction simulations, we explicitly constructed the atomic displacement patterns in the $\mathcal{L}$ and $\mathcal{M}$ phases, which are identical in each layer but differ in their interlayer ordering. Different from the conventional triple-$q$ structure \cite{DiSalvo1976}, the proposed superlattice features a one-dimensional displacement pattern for each of the three types of atoms in the high-temperature unit cell (Ti, Se1, Se2), providing a fresh perspective for investigating the peculiar properties of the CDW phase in 1$T$-TiSe$_2$.

\begin{figure}[tb!]
    \centering
    \includegraphics[width=1\columnwidth]{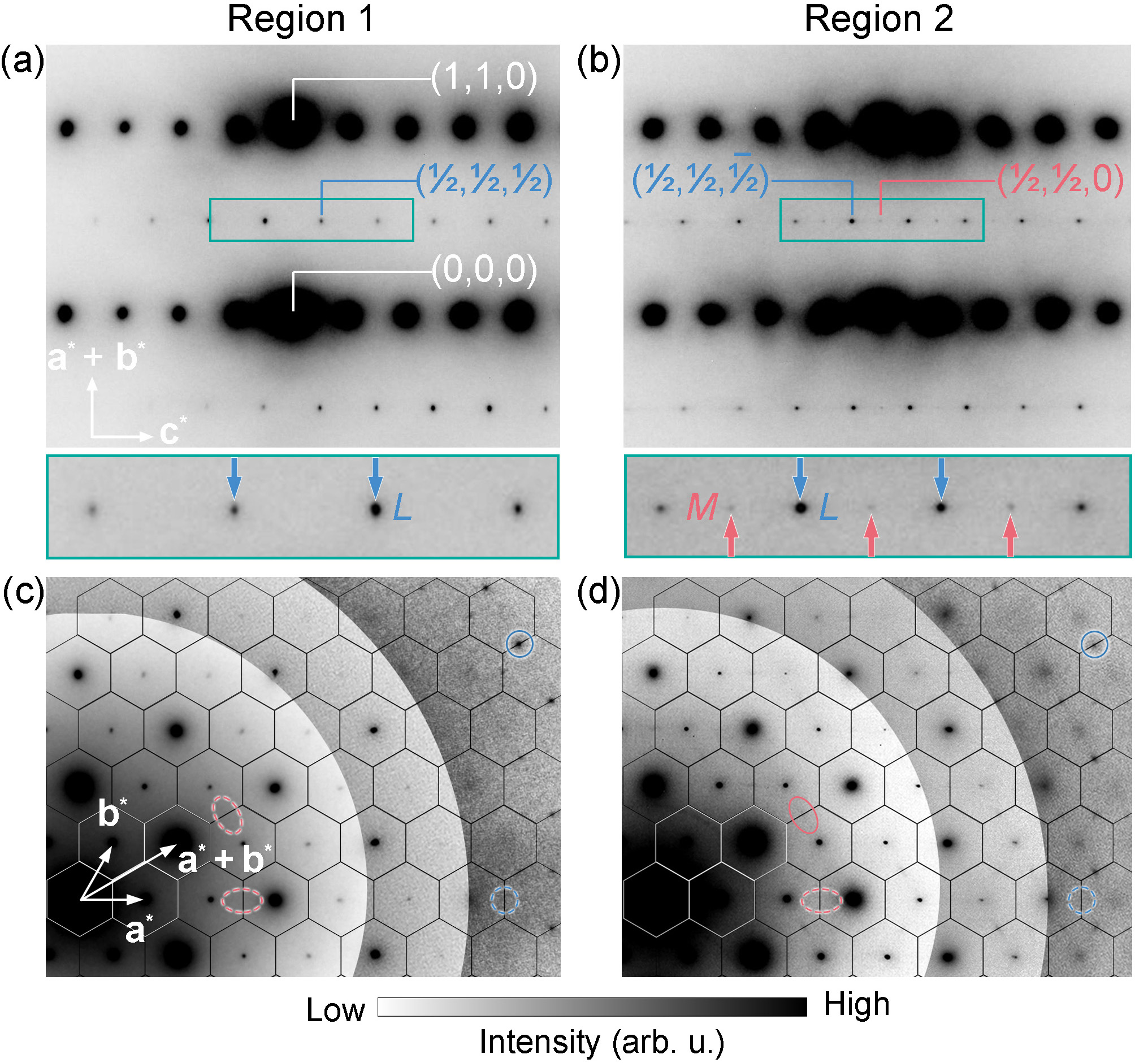}
    \caption{\textbf{Spatially varying diffraction patterns in the CDW phase at 90~K.} (a),(b)~Diffraction patterns along the $[1\overline{1}0]$ zone axis in Regions~1 and 2 of the same sample, respectively. The bottom panels are enlarged views of the respective regions of interest bounded by turquoise rectangles. $L$ peaks (blue arrows) are visible in Region~1, while both $L$ and $M$ peaks (red arrows) are visible in Region~2. (c),(d)~Diffraction patterns along the $[001]$ zone axis in Regions~1 and 2, respectively. The hexagons mark the Brillouin zone boundaries. Solid blue circles (or red ovals) indicate the $L$ (or $M$) peaks, while dashed blue circles (or red ovals) indicate the absence of $L$ (or $M$) peaks. The contrast in the inner and middle circular parts is optimized to emphasize the intensities at $M$ points and Bragg peaks, respectively. The outer edge of the middle circular part indicates the boundary of the zeroth- and half-order Laue zones (see Fig.~S1).}
    \label{fig:Spatially_varying_diffraction_patterns}
\end{figure}

We first investigated the electron diffraction patterns at 90~K, which is much smaller than \TCDW. After scanning different regions of several samples, we observed two qualitatively distinct types of diffraction patterns. The distinction is highlighted in Fig.~\ref{fig:Spatially_varying_diffraction_patterns}(a) and \ref{fig:Spatially_varying_diffraction_patterns}(b), which show cross-sectional diffraction patterns along the $[1\overline{1}0]$ zone axis in Regions~1 and 2, respectively. Here, Miller indices $h$, $k$, and $l$ are assigned according to the high-temperature unit cell [red prism in Fig.~\ref{fig:Real_space_and_momentum_space_unit_cells}(a)]. Diffraction peaks with integral $h$, $k$, and $l$ are Bragg peaks of the parent lattice, while peaks with at least one of $h$ and $k$ being a half-integer are CDW superlattice peaks. A superlattice peak is located at either the $L$ point or the $M$ point of the high-temperature Brillouin zone depending on whether $l$ is a half-integer or an integer, respectively [Fig.~\ref{fig:Real_space_and_momentum_space_unit_cells}(b)]. In Region~1, only $L$ peaks were observed [blue arrows in Fig.~\ref{fig:Spatially_varying_diffraction_patterns}(a)]. On the other hand, in Region~2, sharp peaks at both $L$ and $M$ points were observed [blue and red arrows in Fig.~\ref{fig:Spatially_varying_diffraction_patterns}(b)]. These superlattice peaks persist up to approximately 150~K, above which they become diffuse (Fig.~S3). This contrast between Regions~1 and 2 reveals the mesoscopic spatial heterogeneity of the CDW superlattice that arises from the same homogeneous parent lattice. 

To understand the underlying superlattice structures of Regions~1 and 2 and their connection to reported literature, we searched for published low-temperature structures of 1$T$-TiSe$_2$ based on refinements from X-ray or neutron scatterings \cite{DiSalvo1976, Wegner2020, Kitou2019, Kim2024} and performed electron diffraction simulations (see Fig.~S4 and ref.~\cite{SM} for simulation details). In all cases, we found that $M$ peaks are generally weaker than $L$ peaks but are nevertheless present. Hence, Region~1 without any $M$ peaks observed suggests an unknown CDW structure, which we term the $\mathcal{L}$ phase for its exclusive association with the $L$ peaks (see line profile plots across the $M$ points in Fig.~S2 of ref.~\cite{SM} for a detailed verification of the $M$ peak extinction rules). 

\begin{figure}[tb!]
    \includegraphics[width=\columnwidth]{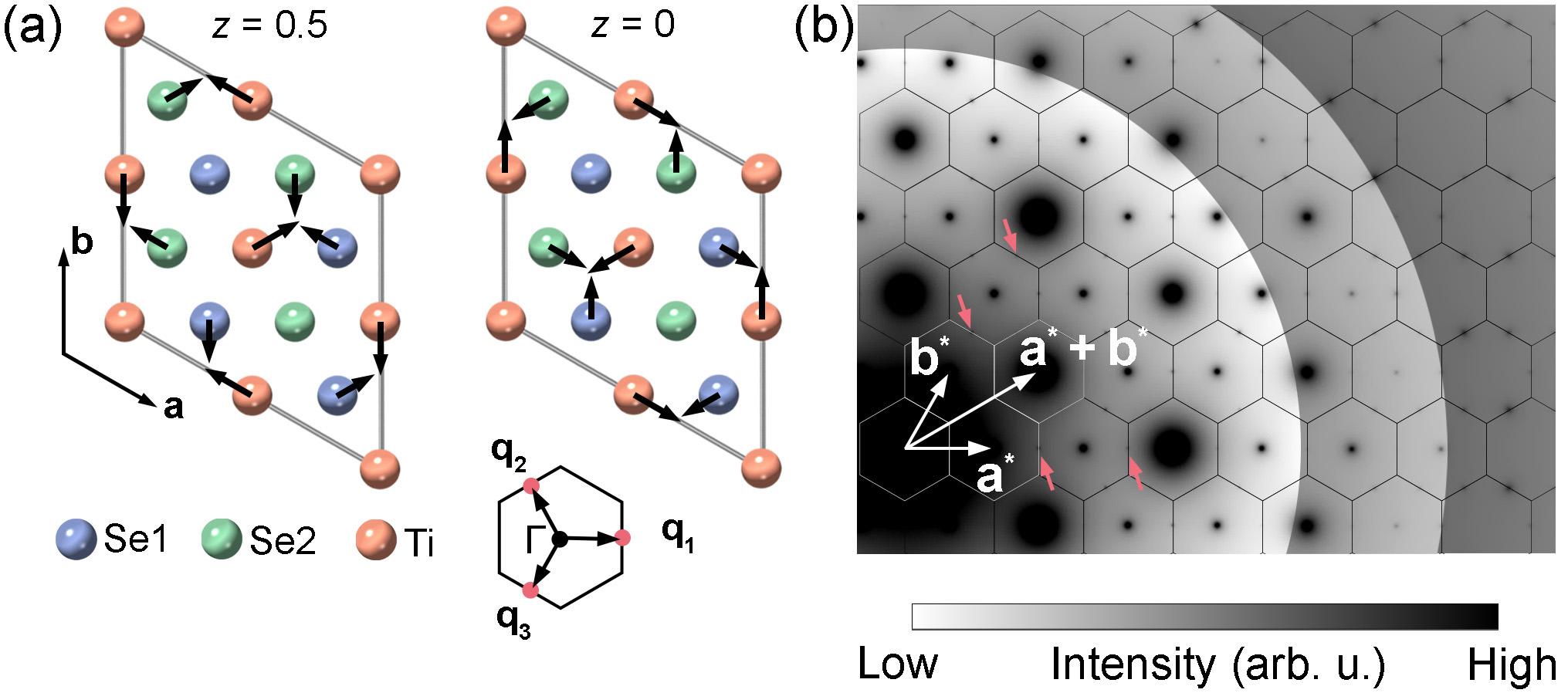}
    \caption{\textbf{Diffraction simulation of the superlattice structure from Di Salvo \textit{et al.}} \cite{DiSalvo1976} (a)~Atomic displacement pattern following the refined CDW superlattice of Di Salvo \textit{et al.}, the prevailing structure in the literature. $z=0$ and 0.5 denote two neighboring layers in the $2\times2\times2$ superlattice, whose displacements are 180$^\circ$ out of phase.  (b)~Simulated electron diffraction pattern along the $[001]$ zone axis based on ref.~\cite{DiSalvo1976}. The contrast in different regions is adjusted following the same procedure as in Fig.~\ref{fig:Spatially_varying_diffraction_patterns}(c)(d). Red arrows point to visible $M$ peaks that are incompatible with observed patterns in either Region~1 or 2 (see Figs.~\ref{fig:Spatially_varying_diffraction_patterns} and S2). 
    Simulated diffraction patterns from other refined superlattice structures in the literature are shown in  Fig.~S4.}
    \label{fig:Simulated_DiSalvo}
\end{figure}

Moreover, Region~2 is also incompatible with the published structures obtained by refinement judging from the observed extinction rules. These rules are clearly seen in the diffraction pattern along the $[001]$ zone axis [Fig.~\ref{fig:Spatially_varying_diffraction_patterns}(d)]. As indicated by the dashed red ovals and blue circles, all $L$ and $M$ peaks with in-plane indices $(h,0)$ and their three-fold symmetry equivalents are extinct. This extinction rule is further confirmed by the cross-sectional diffraction pattern along the $[100]$ zone axis [Fig.~S5(b)], where neither $L$ nor $M$ peaks are visible in any region. By contrast, for the published refined structures, $M$ peaks with in-plane indices $(h,0)$ and their three-fold symmetry equivalents are always present, albeit with varying intensities (Fig.~S4). One such example is highlighted in Fig.~\ref{fig:Simulated_DiSalvo} based on the prevailing superlattice distortion proposed by Di Salvo \textit{et al.} \cite{DiSalvo1976}, which served as the starting point of many other experimental and theoretical analyses of the CDW state \cite{Ishioka2010, Mohr-Vorobeva2011, Castellan2013, Qiao2017, Lin2019, Kim2024, Kim2024b, Cheng2024, Nie2023, Subedi2022}. The simulated pattern based on ref.~\cite{DiSalvo1976} in the $[001]$ zone clearly shows the presence of $M$ peaks with in-plane indices $(h,0)$ and its symmetry-equivalent counterparts [red arrows in Fig.~\ref{fig:Simulated_DiSalvo}(b)]. The disagreement in the extinction rules between our experimental observation [Fig.~\ref{fig:Spatially_varying_diffraction_patterns}(d)] and published data [Figs.~\ref{fig:Simulated_DiSalvo}(b) and S4] hence suggests that Region~2 also hosts a hitherto unknown superlattice structure of 1$T$-TiSe$_2$.

\begin{figure*}[ht]
    \includegraphics[width=1\textwidth]{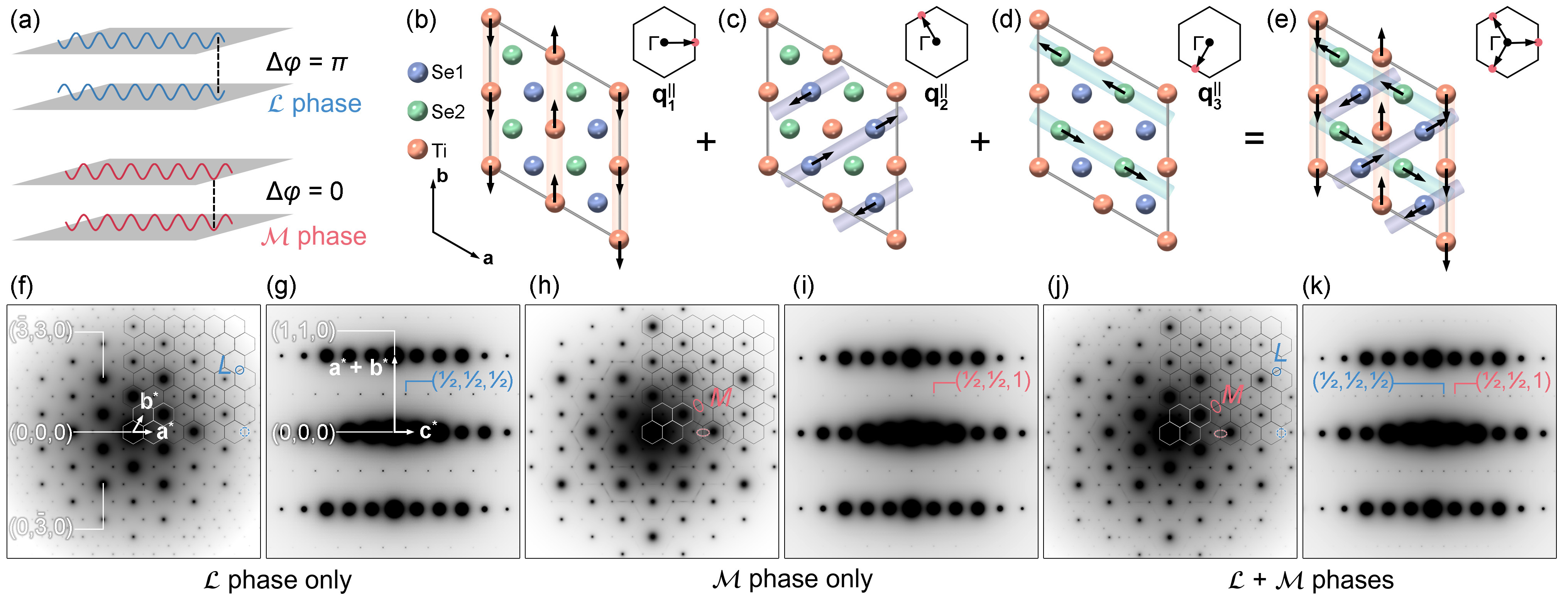}
    \caption{\textbf{One-dimensional atomic displacements and diffraction simulation of the CDW phase.} (a)~Illustration of the interlayer phase difference of atomic displacements in the $\mathcal{L}$ and $\mathcal{M}$ phases. The blue and red waves represent periodically varying atomic displacements in each layer. (b)--(d)~One example of in-plane atomic displacements in either $\mathcal{L}$ or $\mathcal{M}$ phase out of the 216 maximally possible domains compatible with our observed extinction rules, showing a one-dimensional displacement pattern for each atomic type in the parent lattice. In this example, each type of atom Ti, Se1, and Se2 moves in the direction transverse to the in-plane CDW wave vector $\mathbf{q}_1^{\parallel}$, $\mathbf{q}_2^{\parallel}$, and $\mathbf{q}_3^{\parallel}$, respectively. (e)~Overall atomic displacement pattern resulting from the combination of displacements in (b)--(d). (f)--(k)~Simulated electron diffraction patterns along the $[001]$ zone axis [(f),(h),(j)] and the $[1\overline{1}0]$ zone axis [(g),(i),(k)] for the $\mathcal{L}$ phase, the $\mathcal{M}$ phase, and an equal, spatially separated admixture of the two. Solid blue circles (or red ovals) indicate the presence of $L$ (or $M$) peaks, while dashed blue circles (or red ovals) indicate their absence.}
    \label{fig:One-dimensional_displacements_and_simulations}
\end{figure*}

We can gain some insight into the new superlattice structures by noting that the presence of $L$ peaks indicates unit cell doubling along the $c$ axis; on the other hand, if only the $M$ peaks are present, two adjacent layers would have identical atomic displacement patterns. One natural way to realize unit cell doubling along $c$, as suggested by ref.~\cite{DiSalvo1976}, is to have an anti-phase displacement pattern between neighboring layers [Fig.~\ref{fig:Simulated_DiSalvo}(a)]. These considerations suggest that Region~1 with only the $L$ peaks consists of anti-phase CDW layers, whereas Region~2 with both $L$ and $M$ peaks contains an admixture of anti-phase and in-phase CDW layers, as illustrated in Fig.~\ref{fig:One-dimensional_displacements_and_simulations}(a). The coexistence of both types of stacking in the low-temperature state is not unexpected because phonons have been predicted and observed to soften at both $M$ and $L$ points of the Brillouin zone when the material is cooled towards \TCDW \cite{Holt2001,Calandra2011,Otto2021,Fu2016,Cheng2022,Cheng2024}. While the $L$-point soft phonon dominates superlattice formation, fluctuations at the $M$ point can also be locally frozen into a static order, contributing to the observed $M$ peak intensity in Region~2.

To verify whether the $L$ and $M$ peaks in Region~2 indeed represent a mixture of two CDW phases or belong to a single CDW structure, we note that their peak shapes are very different. The $M$ peak is elongated along the direction perpendicular to the in-plane Brillouin zone boundary [solid red oval in Fig.~\ref{fig:Spatially_varying_diffraction_patterns}(d)], but otherwise remains sharp along the other two spatial directions [red arrow in Fig.~\ref{fig:Spatially_varying_diffraction_patterns}(b)]. 
This feature indicates that the spatial distortions responsible for the $M$ peaks have anisotropic correlation lengths. By contrast, $L$ peaks at the same temperature are sharp in all directions [blue arrow in Fig.~\ref{fig:Spatially_varying_diffraction_patterns}(b) and solid blue circle in Fig.~\ref{fig:Spatially_varying_diffraction_patterns}(d)]. Since a single-crystalline CDW superlattice is expected to have one characteristic correlation length, different correlation lengths of the $L$ and $M$ peaks suggest that they indeed originate from separate phases.

The observed phase separation highlights the critical role of spatial resolution in the structural analysis of CDW superlattices. Our selected-area diffraction benefited from a sub-micrometer probe diameter, enabling the isolation of the $\mathcal{L}$ phase that was previously invisible to X-ray or neutron diffraction with a large beam spot, which would average over multiple domains and mix contributions from both $\mathcal{L}$ and $\mathcal{M}$ phases. This spatial averaging likely underlies the discrepancies in the reported low-temperature structures of 1$T$-TiSe$_2$ \cite{DiSalvo1976,Kitou2019,Wegner2020,Kim2024}, as uncontrolled variations in built-in sample strains \cite{Fu2016} or Se vacancies \cite{Huang2017,Campbell2019} could change the $\mathcal{L}$-to-$\mathcal{M}$ phase ratio and hence lead to a different relative intensity between $L$ and $M$ peaks in each of the previously refined structures (Fig.~S4).

To pinpoint the specific atomic displacements in each layer, we analyzed the superlattice extinction rules in the $(h, k)$ indices. These rules are identical for the $L$ and $M$ peaks, suggesting that the $\mathcal{L}$ and $\mathcal{M}$ phases share the same displacement patterns within each layer. A detailed analysis of the extinction rules and scattering structure factors is included in ref.~\cite{SM}; here, we only state the results. Each of the three atomic types (Ti, Se1, Se2) undergoes displacement only along one particular in-plane axis, which is transverse to the respective in-plane component of the CDW wave vectors, $\mathbf{q}^{\parallel}_{1,2,3}$. One example is shown in Fig.~\ref{fig:One-dimensional_displacements_and_simulations}(b)--(d), where Ti, Se1, and Se2 move in three different directions, but only one direction is allowed for each type of atoms. The overall displacement pattern, shown in Fig.~\ref{fig:One-dimensional_displacements_and_simulations}(e), is a combination of the patterns for the three atomic types. To verify our proposed structure, we carried out diffraction simulations along the $[001]$ and $[1\overline{1}0]$ zone axes, assuming the displacements in Fig.~\ref{fig:One-dimensional_displacements_and_simulations}(e) (see ref.~\cite{SM} for simulation details). The simulated patterns for the $\mathcal{L}$ phase [Fig.~\ref{fig:One-dimensional_displacements_and_simulations}(f),(g)] and an equal admixture of spatially separated $\mathcal{L}$ and $\mathcal{M}$ phases [Fig.~\ref{fig:One-dimensional_displacements_and_simulations}(j),(k)] show remarkable resemblance to the measured diffraction patterns in Regions~1 and 2, respectively (Fig.~\ref{fig:Spatially_varying_diffraction_patterns}). In particular, all superlattice extinction rules are obeyed, confirming the consistency between the proposed displacements and experimental observations. 

In contrast to the prevailing triple-$q$ structure \cite{DiSalvo1976}, which retains three-fold rotational symmetry by a microscopic superposition of displacements corresponding to $\mathbf{q}_{1,2,3}$, our proposed structure describes a one-dimensional CDW displacement pattern for each type of atoms. Our structure hence gives a concrete realization of the underlying quasi-one-dimensional CDW chains \cite{VanWezel2010}, as highlighted by the colored stripes in Fig.~\ref{fig:One-dimensional_displacements_and_simulations}(b)--(d). On a general ground, the intra-chain and inter-chain coupling strengths are not expected to be identical, which can naturally lead to in-plane anisotropy, as evidenced by the elongated $M$ peaks seen in the $[001]$ zone [red ovals in Figs.~\ref{fig:Spatially_varying_diffraction_patterns}(d) and \ref{fig:One-dimensional_displacements_and_simulations}(h)] and the reported one-dimensional defect structure after femtosecond photoexcitation \cite{Cheng2024}.

An important implication of our proposed displacement pattern is the presence of a large number of nearly-degenerate structural domains. The displacement rules prescribed by the observed superlattice extinctions allow Ti, Se1, and Se2 to follow one of $\mathbf{q}_1^{\parallel}$, $\mathbf{q}_2^{\parallel}$, or $\mathbf{q}_3^{\parallel}$, so there are $3^3=27$ primary permutations of the wave vectors, as enumerated in Tables~S3--S7. For each wave vector, there are also two equivalent displacement directions that are anti-parallel to each other. In total, there is a maximum of $(3\times 2)^3=216$ possible domains for both $\mathcal{L}$ and $\mathcal{M}$ phases. This large degeneracy can potentially give rise to nano-sized domains within each phase, which are not resolved by the selected-area electron diffraction. The many possible configurations also imply a complex free energy landscape with numerous local minima, making 1$T$-TiSe$_2$ susceptible to optical induction \cite{Xu2020,Jog2023,Wickramaratne2022,Qiu2025} and the formation of metastable states \cite{Duan2023}.

This domain multiplicity also opens up a vast space of unexplored CDW distortions with distinct symmetries. By analyzing the space group assignment of the 27 primary domain configurations within either the $\mathcal{M}$ or $\mathcal{L}$ phase, we identified 18 chiral structures that lack any space-inversion centers, mirror planes, or rotoinversion axes \cite{Wagniere2007} (see Tables~S9 and S10). The proposed structures of 1$T$-TiSe$_2$ as constrained by the observed extinction rules hence provide a new avenue to understand chirality in 1$T$-TiSe$_2$. Our proposal based on measurements of bulk crystals is notably different from previous proposals of chiral structures based on surface-sensitive probes like scanning tunneling microscopy, which rely on a special arrangement of the three CDW wave vectors \cite{Ishioka2010, Kim2024b}. 

In conclusion, our mesoscopic electron diffraction study of 1$T$-TiSe$_2$ uncovers two coexisting CDW phases with distinct interlayer stacking orders, shedding light onto the structural inconsistencies reported in the literature. Contrary to the conventional triple-$q$ model, we identified a one-dimensional displacement pattern for each atomic type, allowing for a multitude of nearly degenerate CDW domains. The proposed structure provides new insights into the metastability, anisotropic in-plane correlations, and the long-debated chiral CDW in this compound. Our findings establish bulk-sensitive mesoscopic probes as essential tools for elucidating complex superlattice structures that can lead to unique states in 1$T$-TiSe$_2$ and other correlated systems both in and out of equilibrium.

\begin{acknowledgments}
We thank helpful discussions with Anshul Kogar and Jasper van Wezel. This work was supported by the U.S. Department of Energy, Office of Basic Energy Science, Division of Materials Science and Engineering, under Contract No. DE-SC0012704. This research used focused ion beam of the Center for Functional Nanomaterials (CFN), which is a U.S. Department of Energy Office of Science User Facility, at Brookhaven National Laboratory under Contract No.~DE-SC0012704.
\end{acknowledgments}

\end{document}


\title{Supplemental Material to \\``Revisiting the charge-density-wave superlattice of 1\!\textit{T}-TiSe$_\text{2}$''}

\author{Wei Wang}
\thanks{These authors contributed equally to this work: W.W. and P.L.}
\affiliation{Condensed Matter Physics and Materials Science Division, Brookhaven National Laboratory, Upton, NY 11973, USA}
\affiliation{Department of Physics, University of Science and Technology of China, Hefei, Anhui 230026, People’s Republic of China}

\author{Patrick Liu}
\thanks{These authors contributed equally to this work: W.W. and P.L.}
\affiliation{Department of Applied Physics, Stanford University, Stanford, California 94305, USA}

\author{Lijun Wu}
\thanks{Correspondence to: ljwu@bnl.gov, alfredz@stanford.edu, and zhu@bnl.gov.}
\affiliation{Condensed Matter Physics and Materials Science Division, Brookhaven National Laboratory, Upton, NY 11973, USA}

\author{Jing Tao}
\affiliation{Condensed Matter Physics and Materials Science Division, Brookhaven National Laboratory, Upton, NY 11973, USA}
\affiliation{Department of Physics, University of Science and Technology of China, Hefei, Anhui 230026, People’s Republic of China}

\author{Genda Gu}
\affiliation{Condensed Matter Physics and Materials Science Division, Brookhaven National Laboratory, Upton, NY 11973, USA}

\author{Alfred Zong}
\thanks{Correspondence to: ljwu@bnl.gov, alfredz@stanford.edu, and zhu@bnl.gov.}
\affiliation{Department of Applied Physics, Stanford University, Stanford, California 94305, USA}
\affiliation{Department of Physics, Stanford University, Stanford, California 94305, USA}
\affiliation{Stanford Institute for Materials and Energy Sciences, SLAC National Accelerator Laboratory, Menlo Park, California 94025, USA}

\author{Yimei Zhu}
\thanks{Correspondence to: ljwu@bnl.gov, alfredz@stanford.edu, and zhu@bnl.gov.}
\affiliation{Condensed Matter Physics and Materials Science Division, Brookhaven National Laboratory, Upton, NY 11973, USA}
\affiliation{Department of Physics and Astronomy, Stony Brook University, Stony Brook, NY 11794, USA}

\date{\today}

\maketitle

\beginsupplement

\tableofcontents

\section{E\lowercase{lectron Diffraction Analysis}}

\subsection{Distinguishing $L$ and $M$ peaks}
\begin{figure}[t]
    \centering
    \includegraphics[width=\columnwidth]{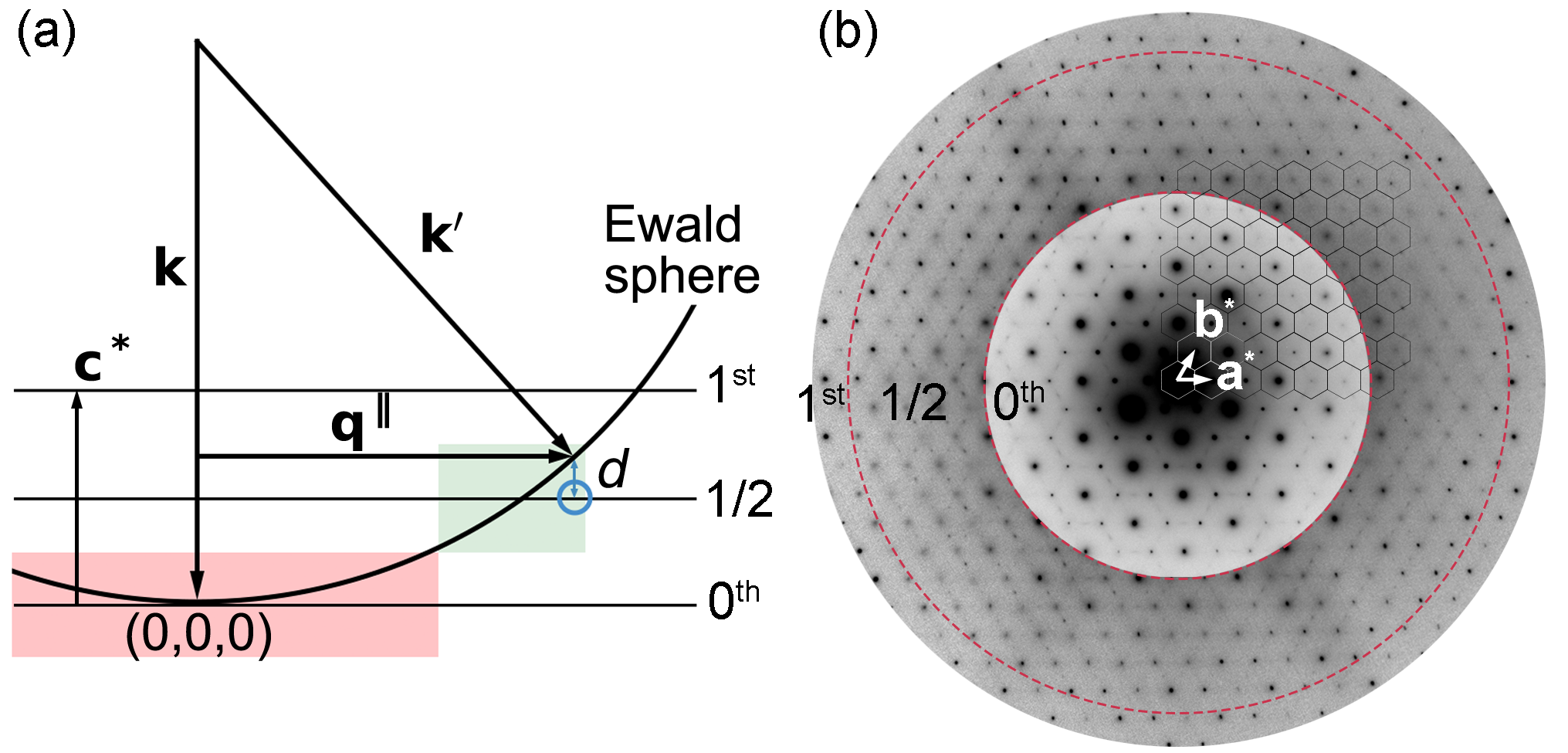}
    \caption{\justifying \textbf{Classification of $L$ and $M$ peaks by the Laue zone order.} (a)~Schematic of the Ewald sphere and Laue zones. $\mathbf{k}$, wave vector of the incident electron beam; $\mathbf{k}'$, wave vector of the scattered beam; $d$, the vertical distance between the Ewald sphere and the set of peaks with Miller index $l$ at the in-plane reciprocal space position $\mathbf{q}^{\parallel}$. The red and green regions label the zeroth- and half-order Laue zones in a diffraction image of the [001] zone axis, respectively. (b)~Diffraction pattern along the $[001]$ zone axis for Region~2, taken at 90~K. The dashed red circles mark the boundaries between the zeroth-, half-, and first-order Laue zones. Image contrasts were adjusted for the zeroth- and half-order Laue zones to emphasize the $M$ peaks and the $L$ peaks, respectively.}
    \label{fig:Ewald_sphere_Laue_zone}
\end{figure}

In this section, we will describe how we discern between $L$ and $M$ superlattice peaks in diffraction patterns along the $[001]$ zone axis. Essentially, superlattice peaks inside the inner red dashed circle in Fig.~\ref{fig:Ewald_sphere_Laue_zone}(b) were identified as $M$ peaks, while those between the inner and outer red dashed circles were identified as $L$ peaks. We will explain these results in the following.

In the diffraction geometry along $[001]$, we define the $n$th-order Laue zone as a region where the Ewald sphere is closest along $c^*$ to the set of peaks with index $l=n$. Here, we used the $l$ Miller index of the high-temperature unit cell, which can be an integer or a half-integer. Correspondingly, $n$ can be an integer or a half-integer as well. Generally, the $l=n$ peaks are the most visible peaks in $n$th-order Laue zone due to their proximity to the Ewald sphere. In Fig.~\ref{fig:Ewald_sphere_Laue_zone}(a), for instance, the $l=0$ peaks are the most visible in the red shaded region, and the $l=1/2$ peaks are the most visible in the green shaded region. 

The \textit{boundary} between the $n$th- and $(n+1/2)$th-order Laue zones in the diffraction pattern is defined as a set of points in the Ewald sphere that is equidistant along $c^*$ from the $l=n$ and $l=n+\frac{1}{2}$ peaks. Based on the geometry in Fig.~\ref{fig:Ewald_sphere_Laue_zone}(a), we can derive the distance $d$ along $c^*$ between the Ewald sphere and the $l=n$ peaks to be
\begin{align*}
d = \left|k - nc^* - \sqrt{{k'}^2 - {q^{\parallel}}^2}\right|,
\tag{S1}
\label{eq:S1}
\end{align*}
where $k=k'\equiv 1/\lambda = 39.87~\text{Å}^{-1}$ are magnitudes of the incident and scattered wave vectors of a 200~keV electron beam, and $q^{\parallel}\equiv|\mathbf{q}^{\parallel}|$ is the magnitude of the in-plane wave vector at which the Ewald sphere is at distance $d$ along $c^*$ from the $l=n$ peaks. 

To distinguish between Laue zones in the diffraction pattern along $[001]$, we note that the vertical distance between the $n$th- and $(n+1/2)$th-order Laue zones is $c^*/2$, so the midpoint is always $d = c^*/4$ along $c^*$ from the $n$th-order Laue zone. Inverting Eq.~\eqref{eq:S1} and solving for $q^{\parallel}$ with $d = c^*/4$ and $n=0$, we calculated the boundary between the zeroth- and half-order Laue zones to be $q^{\parallel} \approx 1.82$~\AA$^{-1}$, which is a circular contour marked by the inner red dashed circle in Fig.~\ref{fig:Ewald_sphere_Laue_zone}(b). Similarly, with $n=1/2$, the boundary between the half- and first-order Laue zones is calculated to be $q^{\parallel} \approx 3.15$~\AA$^{-1}$, as marked by the outer red dashed circle. In the zeroth-order Laue zone ($q^{\parallel}<1.82$~\AA$^{-1}$), $l=n=0$ is integral, so the superlattice peaks here are $M$ peaks. In the half-order Laue zone ($1.82$~\AA$^{-1}<q^{\parallel}<3.15$~\AA$^{-1}$), $l=n=1/2$ is half-integral, so the superlattice peaks here are $L$ peaks.

\subsection{Intensity line profile at $M$ points}
\begin{figure}[t]
    \centering
    \includegraphics[width=1.0\columnwidth]{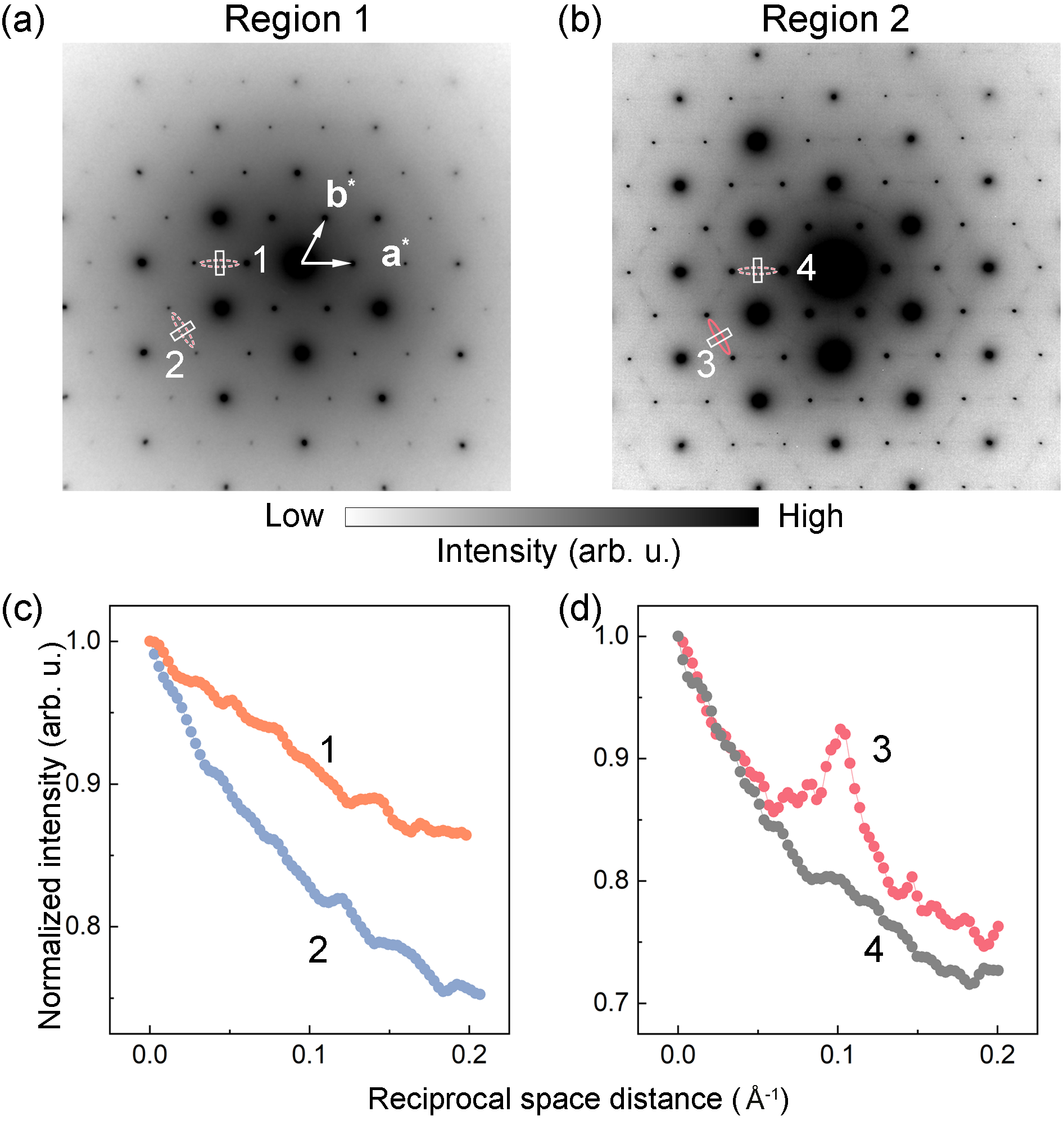}
    \caption{\justifying \textbf{Intensity line profiles near $M$ points in diffraction patterns along the $[001]$ zone axis}. (a),(b)~Electron diffraction patterns from Regions~1 and 2. The dashed (solid) red ovals represent the absence (presence) of $M$ peaks at reciprocal space positions. White rectangles perpendicular to the red ovals indicate four regions of interest (ROIs) labeled~1--4. Line profiles are extracted along the long axis of the ROIs, where the $M$ peaks are located at 0.1~\AA. (c),(d)~Intensity line profiles along (c) ROIs~1,~2 and (d) ROIs~3,~4. The changing background arises from the slightly unequal intensities of the two Bragg peaks on the two sides of the long axis of the ROIs. In Region~1, peaks at the $M$ point are absent in both cuts~1 and 2. In Region~2, peaks at the $M$ point are present in cut~3 but absent in cut~4.}
    \label{fig:M_linecuts}
\end{figure}

To confirm the $M$ peak extinction rules observed in diffraction patterns along the $[001]$ zone axis [Fig.~2(c),(d)], we measured the line profiles cutting through $M$ points along \( \mathbf{a}^* \) and \( \mathbf{a}^* + \mathbf{b}^* \) in the diffraction patterns from Regions~1 [Fig.~\ref{fig:M_linecuts}(a)] and 2 [Fig.~\ref{fig:M_linecuts}(b)], which are the same as Fig.~2(c) and 2(d). Here, solid red oval labels the presence of $M$ peaks, and dashed red ovals label the absence of $M$ peaks.  Since the $M$ points are located midway between two Bragg peaks, we first extracted a line profile spanning two Bragg peaks and identified their midpoint as the $M$ point. We then extracted the intensity line profiles in regions of interest close to $M$ points along \( \mathbf{a}^* \) and \( \mathbf{a}^* + \mathbf{b}^* \) directions, as indicated by the white rectangles in Fig.~\ref{fig:M_linecuts}. The normalized intensities measured along the white rectangles 1, 2 and 3, 4 are plotted in Fig.~\ref{fig:M_linecuts}(c) and Fig.~\ref{fig:M_linecuts}(d), respectively.  Our analysis reveals that no $M$ peaks along either the \( \mathbf{a}^* \) or \( \mathbf{a}^* + \mathbf{b}^* \) directions are observable in Fig.~\ref{fig:M_linecuts}(a), demonstrating the extinction of $M$ peaks in Region~1. $M$ peaks are extinct along the \( \mathbf{a}^* \) direction but present along \( \mathbf{a}^* + \mathbf{b}^* \) in Fig.~\ref{fig:M_linecuts}(b), supporting the extinction rules we proposed for Region~2. In other words, $M$ peaks are present along \( \mathbf{a}^* + \mathbf{b}^* \) in some regions of the sample but absent along  \( \mathbf{a}^* \) throughout the sample. The same conclusion holds for all directions that are symmetry equivalents of \( \mathbf{a}^* \) and \( \mathbf{a}^* + \mathbf{b}^* \).

\begin{figure}[ht]
    \centering
    \includegraphics[width=0.8\columnwidth]{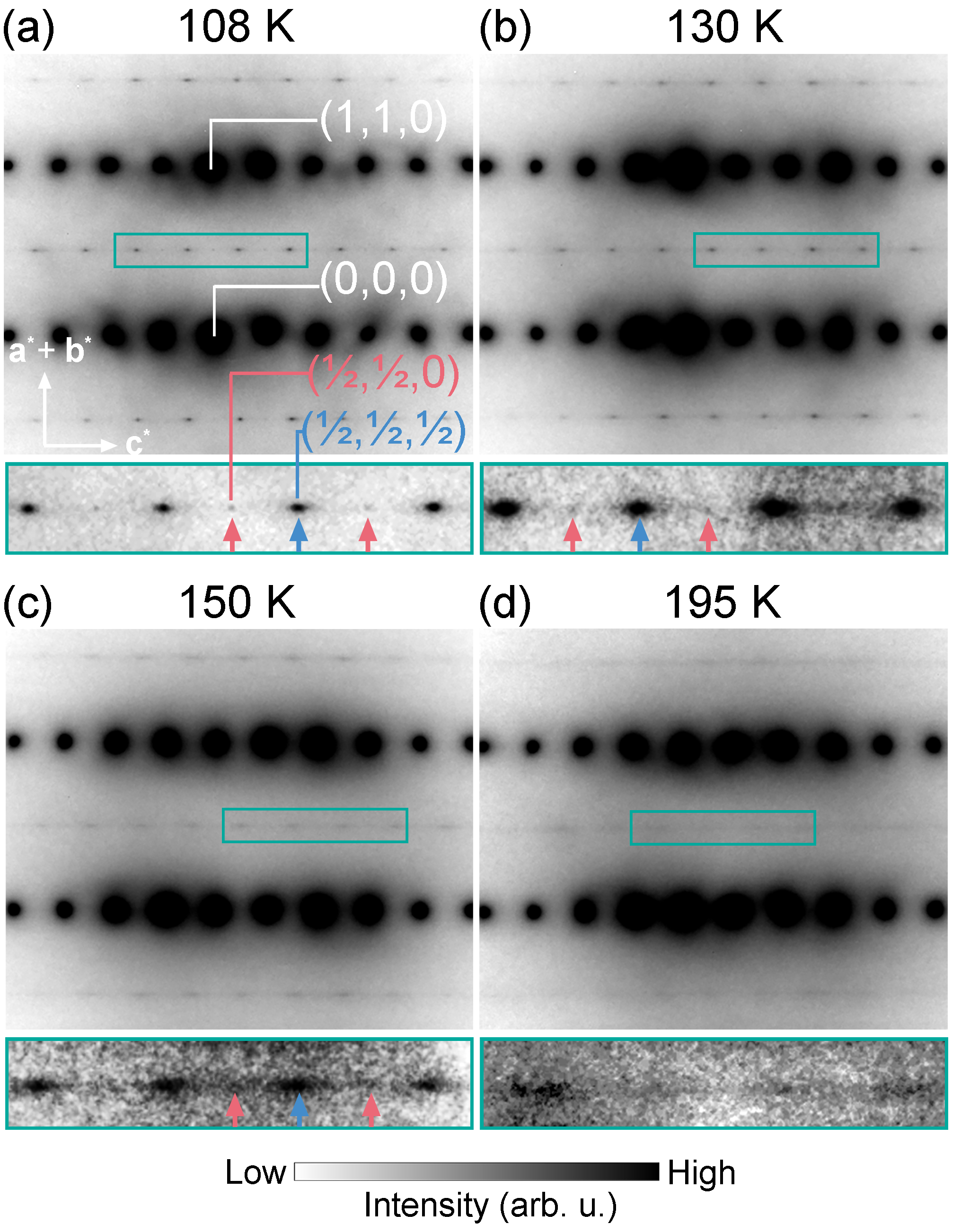}
    \caption{\justifying \textbf{Temperature evolution of the \textit{L} and \textit{M} superlattice peaks in Region~2.} (a)--(d)~Diffraction patterns in the $[1\overline{1}0]$ zone at four temperatures below \TCDW, with blue (or red) arrows indicating the $L$ (or $M$) peaks. The patterns are presented on a logarithmic scale to enhance clarity and emphasize the superlattice peaks. Bottom insets are enlarged views of the $(\frac{1}{2},\frac{1}{2},l)$ row with increased contrast to better visualize the $L$ and $M$ peaks.}
    \label{fig:T_evolution_L_and_M_peaks}
\end{figure}

\subsection{Temperature evolution of superlattice peaks}

We measured the cross-sectional diffraction pattern along the $[1\overline{1}0]$ zone axis at various temperatures (Fig.~\ref{fig:T_evolution_L_and_M_peaks}). The $L$ and $M$ superlattice peaks remain sharp and distinguishable until approximately 150~K, above which diffuse intensities along the $\mathbf{c}^*$ direction are significantly enhanced.

\section{M\lowercase{aterials and methods}}

\subsection{Sample preparation}
1$T$-TiSe$_2$ single crystals were grown using a self-flux method. A mixture of 2\% high-purity (99.999\%) Ti and 98\% high-purity (99.9999\%) Se was sealed in a quartz tube. The tube was heated to 1000$^{\circ}$C and maintained at that temperature in a box furnace for two days before being cooled to 250$^{\circ}$C over 200~hours. It was then reheated to 400$^{\circ}$C within 5~minutes, after which the liquid selenium in the tube was poured into one end, while the single crystals remained at the other end. Finally, the tube was cooled to room temperature over 12~hours. 

For transmission electron microscope (TEM) measurements, $[001]$-oriented samples were prepared using mechanical exfoliation with Scotch tapes. The bulk material was first affixed to a glass slide substrate using Crystal Bond, followed by iterative peeling to progressively thin the sample. The sample thickness was estimated via an optical microscope by analyzing color contrast variations. Once the samples reached a thickness sufficient for electron transparency, the samples were detached from the substrates by floating in acetone and scooped onto standard 300-mesh copper TEM grids. The lateral dimensions of the exfoliated samples were approximately $100~\upmu\mathrm{m}$, and sample thickness was approximately 30--40~nm. This sample preparation procedure is similar to the methodology outlined in ref.~\cite{Cheng2024}. Cross-sectional $[100]$ and $[1\overline{1}0]$-oriented samples were prepared using Thermo Fisher Scientific Helios G5 Dual Beam Focused Ion Beam (FIB) Microscope with Ga ion. The lateral dimensions of the FIB samples were approximately $10\, \upmu\mathrm{m}$ by $6\,\upmu\mathrm{m}$, with a thickness of around 60--70\,\text{nm}. The thickness of FIB samples can be estimated by the sample and vacuum background contrast in the electron-beam image in FIB during the sample preparation. 

\subsection{Electron diffraction measurements}
Diffraction patterns were collected using a JEOL ARM-200F microscope operating at 200~kV, equipped with a Gatan Orius SC200D CCD camera, located at Brookhaven National Laboratory. TEM experiments were conducted with a Gatan liquid nitrogen cooling stage, featuring double tilt capability, with the lowest attainable temperature reaching approximately 90~K. The estimated total electron counts are approximately $1.3 \times 10^{9}~\text{e}^-$ in the diffraction patterns along $[001]$, e.g., Fig.~\ref{fig:Ewald_sphere_Laue_zone}(b), and approximately $8.6 \times 10^{8}~\text{e}^-$ in the diffraction patterns along $[1\overline{1}0]$, e.g., Fig.~\ref{fig:T_evolution_L_and_M_peaks}(a)--(d). The selected-area aperture used in the experiment had a diameter of 240~nm.

\subsection{Electron diffraction simulation}
The dynamic diffraction simulation is based on the Bloch wave method, implemented with the computer codes developed in-house at Brookhaven National Laboratory. Simulations were conducted at an accelerating voltage of 200~kV (de~Broglie wavelength = 0.02507~\AA) for comparison with TEM experimental data. The sample thickness range was chosen to match the intensity distribution of multiple Bragg peaks. Given local thickness variations of the experimental samples, we tested $[001]$-oriented samples from 30--39~nm and cross-sectional $[100]$- and $[1\overline{1}0]$-oriented samples from 60--69~nm, both with a 1-nm step size in the simulation. Simulated diffraction patterns varied minimally within each range, so the final results represent an average over these thickness values.

\section{D\lowercase{iffraction pattern simulations for published structures}}

Various published refined structures proposed inconsistent space groups for the CDW superlattice of 1$T$-TiSe$_2$, as summarized in Table~\ref{tab:published_structures} based on refs.~\cite{DiSalvo1976, Wegner2020, Kitou2019, Kim2024}. To examine the similarities and differences between the published structures and our experimental data, we performed dynamic electron diffraction simulations using the published structures. The results are summarized in Fig.~\ref{fig:Simulated_published_structures}. The first column of Fig.~\ref{fig:Simulated_published_structures}(a)--(f) are simulated diffraction patterns along the $[001]$ zone axis. $M$ peaks with indices $(h,0,l)$ and their three-fold symmetry equivalents are present, as highlighted by red arrows in the insets. These peaks are not observed in the experimental diffraction pattern along the $[001]$ zone axis. This inconsistency is further evidenced by comparing the second column of Fig.~\ref{fig:Simulated_published_structures}(a)--(f), simulated diffraction patterns for published structures along the $[100]$ zone axis, and Fig.~\ref{fig:diffraction_along_100}(b), the experimental cross-sectional diffraction pattern along the $[100]$ zone axis. While $M$ peaks are present in Fig.~\ref{fig:Simulated_published_structures}(a)--(f), they are extinct in Fig.~\ref{fig:diffraction_along_100}(b). Note that the $M$ peaks in this cross section have indices $(0,k,l)$, which are three-fold symmetry equivalents of $(h,0,l)$ and obey the same extinction rules. The third column of Fig.~\ref{fig:Simulated_published_structures}(a)--(f) presents simulated diffraction patterns along the $[1\overline{1}0]$ zone axis. Both the $L$ and $M$ peaks with indices $(h,h,l)$ are present in the simulated patterns in Fig.~\ref{fig:Simulated_published_structures}(e),(f). The presence of $M$ peaks in the simulated patterns is inconsistent with Region~1 of the experimental data where only $L$ peaks with indices $(h,h,l)$ are present [Fig.~2(a)]. These discrepancies suggest that the proposed structural models cannot fully reproduce the experimentally observed diffraction patterns of the 1$T$-TiSe$_2$ CDW superlattice. As a comparison, we also performed kinematic simulations. The results are qualitatively consistent with the dynamic simulation results, confirming that the superlattice extinction rules in the dynamic simulations are not artifacts of multiple scattering.

\begin{figure*}[!htbp]
      \centering
      \includegraphics[width=0.95\textwidth]{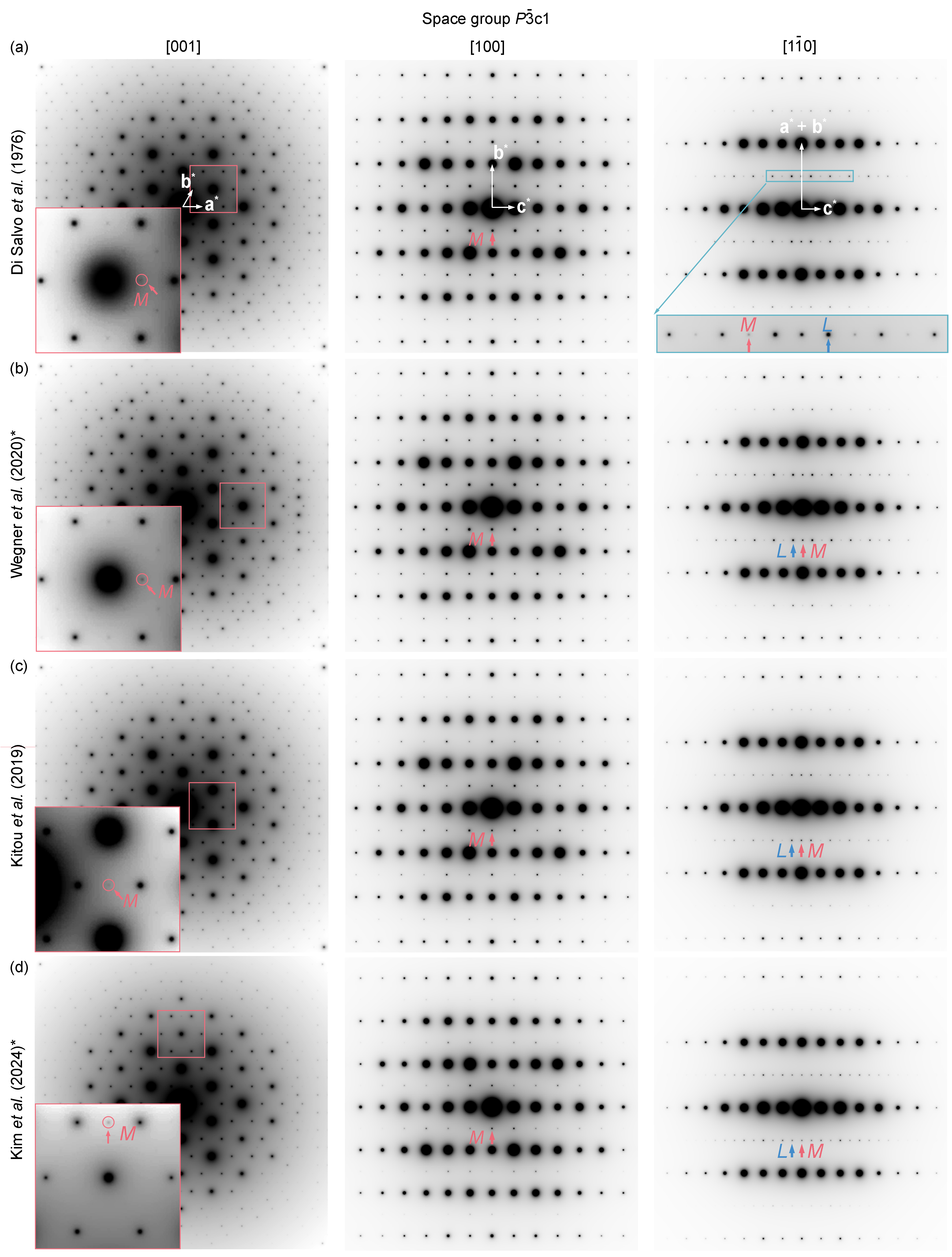}
      \label{fig:subfig1}
\end{figure*}

\begin{figure*}[!htbp]
    \centering
    \includegraphics[width=0.95\textwidth]{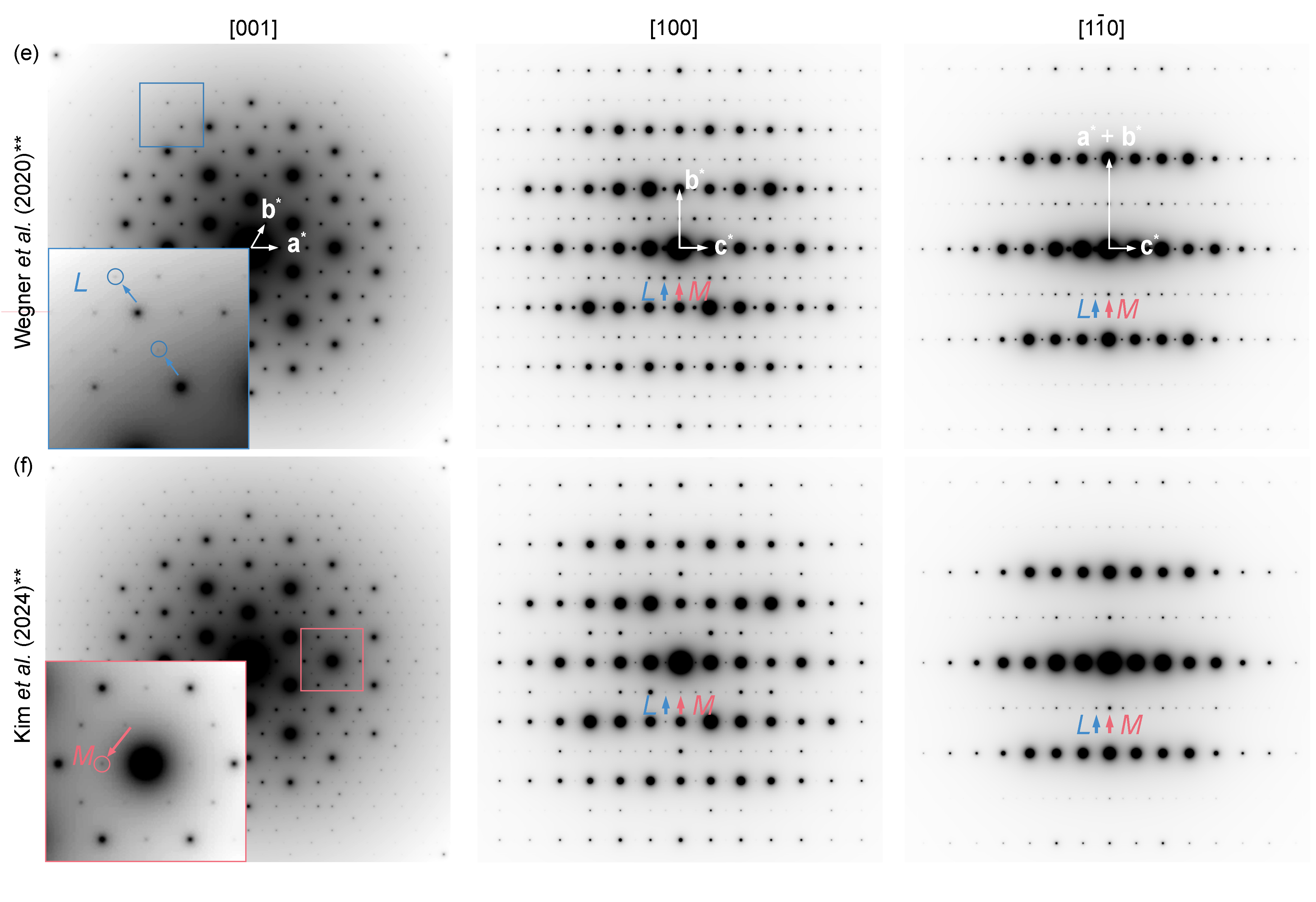}
    \label{fig:subfig2}
    
    \caption{\justifying \textbf{Electron diffraction patterns simulated using published refined structures.} Dynamic diffraction simulations were carried out along $[001]$, $[100]$, and $[1\overline{1}0]$ zone axes using published structures listed in Table~\ref{tab:published_structures}. The red arrows indicate the $M$ peaks, and the blue arrows indicate the $L$ peaks. The space group of the structural model is $P\overline{3}c1$ in (a)--(d), $P\overline{3}m1$ in (e), and $P321$ in (f). The structure in (b) was a reinterpretation by Wegner \textit{et al.} of the structure from Di~Salvo \textit{et al.}; on the other hand, the structure in (e) denotes a pseudo-Jahn-Teller distortion refined independently by Wegner \textit{et al.} \cite{Wegner2020}. The structure in (d) has a space group $P\overline{3}c1$ while that in (f) has a space group $P321$, both of which were compatible with the refinement results of Kim \textit{et al.} \cite{Kim2024}. Insets of figures in the first column are enlarged views of the boxed regions in the corresponding panels. The enlarged areas are selected differently in each panel to optimize the visibility of superlattice peaks, which varies across different structural models. The extracted extinction rules are the same due to the three-fold symmetry of the parent Brillouin zone.}
    \label{fig:Simulated_published_structures}
\end{figure*}

\afterpage{
  \begin{table*}[!htbp]
    \centering
    \renewcommand{\arraystretch}{1.1}   
    \caption{Published refined structures of 1$T$-TiSe$_2$.}
    \begin{tabular}{c|c|c}
        \Xhline{0.25ex}
        \textbf{Publication} & \textbf{Space group} & \textbf{Summary of structure}\\
        \hline
        Di Salvo \textit{et al.}, \textit{Phys. Rev. B} \textbf{14}, 4321 (1976) \cite{DiSalvo1976}& \multirow{4}{*}{\makecell{\\ \\ \\$P\overline{3}c1$} } & Triple-\textit{q} displacements with interlayer anti-phase stacking\\
        \cline{1-1} \cline{3-3}
        Wegner \textit{et al.}, \textit{Phys. Rev. B} \textbf{101}, 195145 (2020) \cite{Wegner2020}&  & \makecell{Di Salvo \textit{et al.} structure interpreted by Wegner \textit{et al.}, \\ but lacks interlayer anti-phase stacking}\\
        \cline{1-1} \cline{3-3}
        Kitou \textit{et al.}, \textit{Phys. Rev. B} \textbf{99}, 104109 (2019) \cite{Kitou2019} &  & \makecell{Similar to Di Salvo \textit{et al.}, but lacks interlayer \\ anti-phase stacking }\\
        \cline{1-1} \cline{3-3}
        Kim \textit{et al.}, \textit{Nat. Phys.} \textbf{20}, 1919--1926 (2024) \cite{Kim2024}& & \makecell{Similar to Di Salvo \textit{et al.}, but lacks interlayer \\ anti-phase stacking }\\ 
        \hline
        Wegner \textit{et al.}, \textit{Phys. Rev. B} \textbf{101}, 195145 (2020) \cite{Wegner2020} & $P\overline{3}m1$ & Breathing-type pseudo-Jahn-Teller distortion \\
        \hline
        Kim \textit{et al.}, \textit{Nat. Phys.} \textbf{20}, 1919--1926 (2024) \cite{Kim2024} & $P321$  & \makecell{Chiral structure from the intersection of electronic and \\ lattice space groups}\\
        \Xhline{0.25ex}
    \end{tabular}
    \label{tab:published_structures}
  \end{table*}
}

\begin{figure}[h]
    \centering
    \includegraphics[width=\columnwidth]{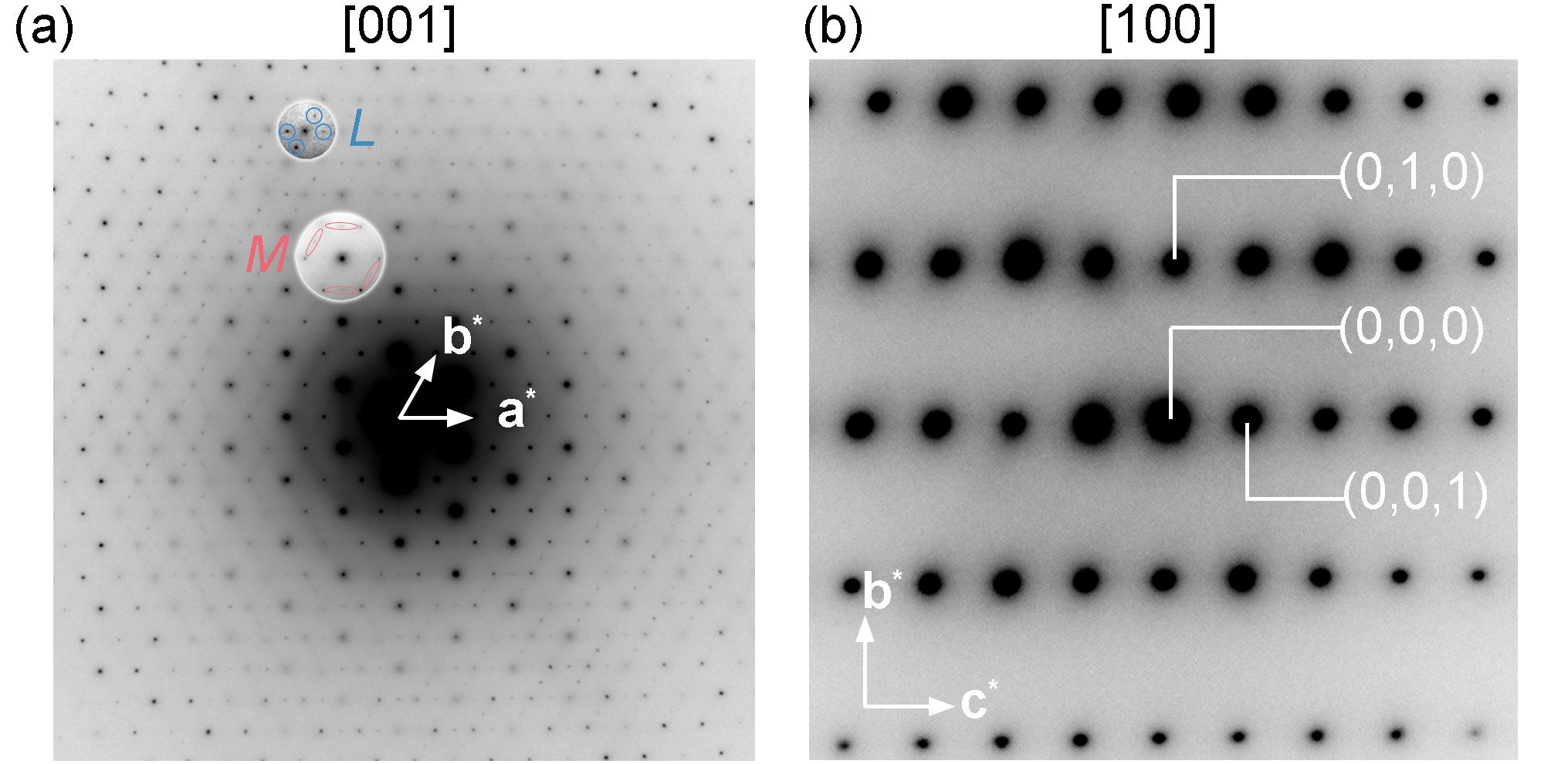}
    \caption{\justifying \textbf{Experimental electron diffraction patterns taken at 90 K}. Measurements were conducted along (a)~the $[001]$ zone axis and (b)~the $[100]$ zone axis. The contrasts of selected areas are adjusted to highlight the superlattice peaks. Red ovals and blue circles indicate the $M$ and $L$ peaks, respectively. }
    \label{fig:diffraction_along_100}
\end{figure}

\section{S\lowercase{tructural Analysis of the} $\mathcal{L}$\lowercase{ Phase}}

The following two sections aim at uncovering the different atomic displacement configurations of the $\mathcal{L}$ and $\mathcal{M}$ phases that are responsible for their superlattice extinction rules observed in our diffraction patterns.

\begin{figure}[h]
    \centering
    \includegraphics[width=\columnwidth]{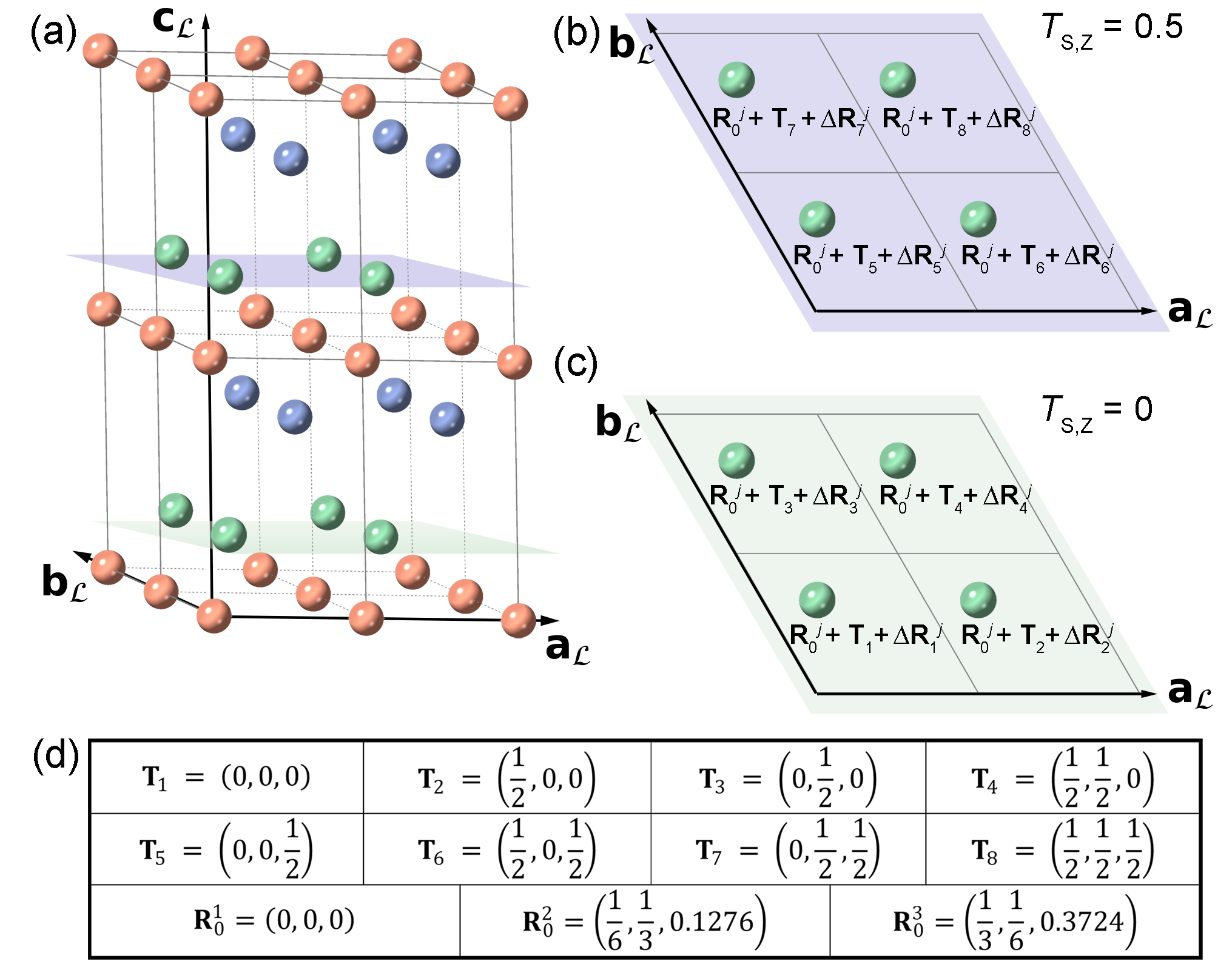}
    \caption{\justifying \textbf{Structure of the $2 \times 2 \times 2$ $\mathcal{L}$ phase superlattice.} (a) 3D schematic of the superlattice structure. (b) Schematic of top-layer subunit cells $(s = 5,6,7,8)$ with Se1 shown; (c) Schematic of bottom-layer subunit cells $(s=1,2,3,4)$ with Se1 shown. (d) A list of values of $\mathbf{T}_s$, the translation vectors relating the first subunit cell to the $s$th subunit cell, and $\mathbf{R}_0^j$, the position of the $j$th atom in the high-temperature phase in the first subunit cell. $j = 1, 2, 3$ for Ti, Se1, and Se2, respectively. All coordinates are written as a fraction of the $2 \times 2 \times 2$ lattice vectors. See text for the definition of $\Delta\mathbf{R}_s^j$.}
    \label{fig:L_supercell}
\end{figure}

The high-temperature 1$T$-TiSe\textsubscript{2} unit cell consists of three atomic types: Ti, Se1, and Se2. In this section, we first construct the \(2 \times 2 \times 2\) superlattice of the $\mathcal{L}$ phase, consisting of eight high-temperature unit cells (referred to as subunit cells) [Fig.~\ref{fig:L_supercell}(a)--(c)]. The position of an atom within a superlattice cell is given by
\begin{align*}
\mathbf{R}_s^j = \mathbf{R}_0^j + \mathbf{T}_s + \Delta \mathbf{R}_s^j.
\tag{S2}
\label{eq:S2}
\end{align*}
Here, $j = 1, 2, 3$ enumerates the atoms in one subunit cell, Ti, Se1, and Se2. The labels \(s = 1, 2, 3, 4\) and \(s = 5, 6, 7, 8\) indicate the subunit cells in the bottom and top layers of the supercell, respectively. \(\mathbf{R}_0^j\) is the fractional position of the \(j\)th atom in the first subunit cell of the \(2 \times 2 \times 2\) supercell before the CDW displacement. Explicitly, $\mathbf{R}_0^j = \mathbf{r}_0^j/2$, where $\mathbf{r}_0^j$ is the fractional position of the $j$th atom in the $1 \times 1 \times 1$ high-temperature unit cell. $\mathbf{T}_s$ is the translation vector that relates the $s$th subunit cell to the first, such that $\mathbf{R}_0^j+\mathbf{T}_s$ represents the position of the $j$th atom in the $s$th supercell before the displacement. The values of each $\mathbf{R}_0^j$ and $\mathbf{T}_s$ are listed in Fig.~\ref{fig:L_supercell}(d). $\Delta \mathbf{R}_s^j$ denotes the CDW displacement vector of the $j$th atom in the $s$th supercell. Finally, $\mathbf{R}_s^j$ is the position of the $j$th atom in the $s$th subunit cell after the displacement. 

\begin{figure}[h]
    \centering
    \includegraphics[width=\columnwidth]{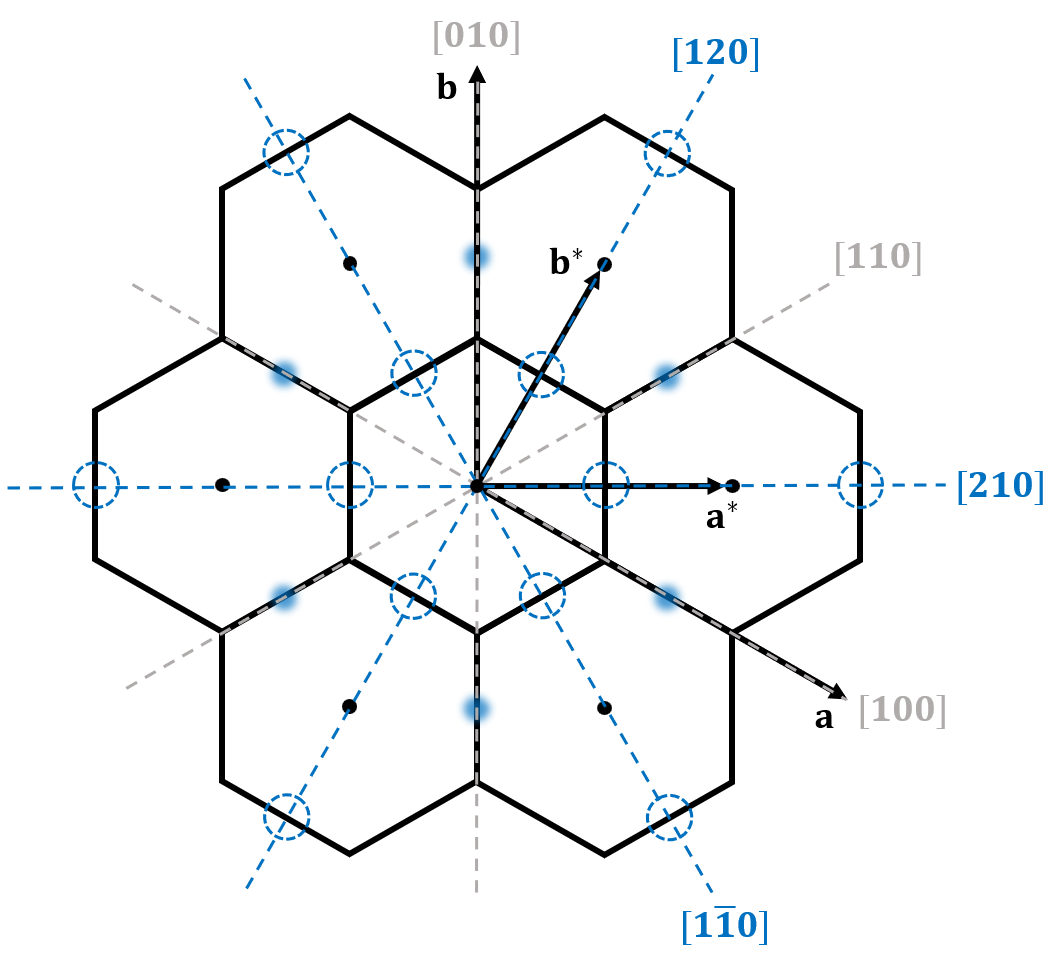}    \caption{\justifying\textbf{$L$ peak extinction rules in the $\mathcal{L}$ phase.} Blue dash lines represent zone axes along which $L$ peaks are visible in diffraction patterns, while gray dash lines represent those along which the $L$ peaks are extinct. The blue disks mark the in-plane projection of visible $L$ peaks, while the dashed blue circles mark the $L$ points where the peak is extinct.}
    \label{fig:L_extinction_rules}
\end{figure}

In this section, we analyze the CDW atomic displacements associated with the $\mathcal{L}$ phase. Table \ref{tab:extinction_rules} summarizes the experimental observations of $L$ peaks in diffraction patterns along each zone axis. The corresponding picture is shown in Fig. \ref{fig:L_extinction_rules}: for diffraction patterns taken along the zone axes marked by blue dashed lines, the $L$ peaks are present; for diffraction patterns taken along the gray dashed lines, the $L$ peaks are extinct. $M$ peaks are fully extinct along any zone axis for this phase. To better understand the implications of these observations, we analytically calculated the superlattice extinction rules for some atomic displacement configurations. Displacement configurations that allow us to reproduce the observed extinction rules would be considered as a more accurate representation of the CDW phases of 1$T$-TiSe$_2$.

\begin{table}[h]
    \centering
    \caption{\justifying\textbf{Present and extinct $L$ peaks in diffraction patterns along various zone axes.} Blue text indicates zone axes along which $L$ peaks are present; gray text indicates zone axes along which $L$ peaks are extinct.}
    \begin{tabular}{c|c}
    \Xhline{0.25ex}
    \textbf{Zone axis} & \textbf{Present $L$ peaks} \\
    \hline
    \textcolor{NavyBlue}{$[1\overline{1}0]\quad$} & \textcolor{NavyBlue}{$(h, h, l)$} \\
    \textcolor{NavyBlue}{$[120]\quad$} & $\textcolor{NavyBlue}{(2h, \overline{h}, l)}$ \\
    \textcolor{NavyBlue}{$[210]\quad$} & $\textcolor{NavyBlue}{(h, \overline{2h}, l)}$ \\
    \Xhline{0.25ex}
    \textbf{ } & \textbf{Extinct $L$ peaks} \\
    \hline
    \textcolor{gray}{$[010]\quad$} & \textcolor{gray}{$(h, 0, l)$} \\
    \textcolor{gray}{$[100]\quad$} & \textcolor{gray}{$(0, k, l)$} \\
    \textcolor{gray}{$[110]\quad$} & \textcolor{gray}{$(h, \overline{h}, l)$} \\
    \Xhline{0.25ex}
    \end{tabular}
    \label{tab:extinction_rules}
\end{table}

First, we write down the expression for the superlattice structure factor. We start with the structure factor as a function of the reciprocal lattice vector $\mathbf{G}$,
\begin{align*}
F_{\mathbf{G}} 
&= \sum_{j=1}^{3} \sum_{s=1}^{8} f^{j} e^{i2\pi\mathbf{G}\cdot\bigl(\mathbf{R}_{0}^{j} + \mathbf{T}_{s} + \Delta \mathbf{R}_{s}^{j}\bigr)} \\
&= \sum_{j=1}^{3} f^{j} F_{\mathbf{G}}^{j} e^{i2\pi\left(\mathbf{G}\cdot \mathbf{R}_{0}^{j}\right)},
\tag{S3}
\label{eq:S3}
\end{align*}
where $f^{j}$ is the atomic form factor of atom $j$. $F_{\mathbf{G}}^{j}$ is the structure factor for the atomic type $j$, defined as

\begin{align*}
F_{\mathbf{G}}^{j}
=& \sum_{s=1}^{8} e^{i2\pi\mathbf{G}\cdot\left(\mathbf{T}_{s} + \Delta \mathbf{R}_{s}^{j}\right)} \\
=& \bigl[e^{i2\pi\mathbf{G}\cdot\Delta \mathbf{R}_{1}^{j}} + e^{i2\pi l}e^{i2\pi\mathbf{G}\cdot\Delta \mathbf{R}_{5}^{j}}\bigr] \\
& + e^{i2\pi h}\bigl[e^{i2\pi\mathbf{G}\cdot\Delta \mathbf{R}_{2}^{j}} + e^{i2\pi l}e^{i2\pi\mathbf{G}\cdot\Delta \mathbf{R}_{6}^{j}}\bigr] \quad \\
&+ e^{i2\pi k}\bigl[e^{i2\pi\mathbf{G}\cdot\Delta \mathbf{R}_{3}^{j}} + e^{i2\pi l}e^{i2\pi\mathbf{G}\cdot\Delta \mathbf{R}_{7}^{j}}\bigr]\\
&+e^{i2\pi(h+k)}\bigl[e^{i2\pi\mathbf{G}\cdot\Delta \mathbf{R}_{4}^{j}} + e^{i2\pi l}e^{i2\pi\mathbf{G}\cdot\Delta \mathbf{R}_{8}^{j}}\bigr],
\tag{S4}
\label{eq:S4}
\end{align*}

\noindent
where $\mathbf{G} \equiv 2h\mathbf{a}_{\mathcal{L}}^{\ast} + 2k\mathbf{b}_{\mathcal{L}}^{\ast} + 2l\mathbf{c}_{\mathcal{L}}^{\ast},$ and $\mathbf{a}_{\mathcal{L}}^{\ast}$, $\mathbf{b}_{\mathcal{L}}^{\ast}$, $\mathbf{c}_{\mathcal{L}}^{\ast}$ are the reciprocal lattice vectors for the $2\times2\times2$ supercell of the $\mathcal{L}$ phase, and $(h, k, l)$ are the Miller indices of the high-temperature unit cell.

For a peak at $(h, k, l)$ to be a superlattice peak, at least one of $h$, $k$, or $l$ must be a half-integer. Under this condition, the factors $e^{i2\pi l}$, $e^{i2\pi h}$, $e^{i2\pi k}$, and $e^{i2\pi(h+k)}$ are either $+1$ or $-1$, depending on whether $h$, $k$, and $l$ are integers or half-integers. 

Now let us consider the atomic displacement configurations that can reproduce the observed $M$ and $L$ peak extinction rules when they are applied to Eq.~\eqref{eq:S4}. A summary of the various conditions on atomic displacement vectors, their corresponding physical schematics, and the resulting superlattice extinction rules are presented in Table~\ref{tab:L_phase_conditions}. In the following, we will derive the rules in this table.

First, we need to ensure that all $M$ peaks in the $\mathcal{L}$ phase are strictly forbidden. As explained in the main text, we take the interlayer anti-phase stacking order of 1$T$-TiSe$_2$ as the starting point. Let us call it \textbf{Condition~1} (see Table~\ref{tab:L_phase_conditions}). This condition generally leads to weak $M$ peaks compared to the $L$ peaks, but it is not sufficient to ensure strictly extinct $M$ peaks in the $\mathcal{L}$ phase. To explore the additional conditions needed, let us first compute $F_{\mathbf{G}}^j$ for the $M$ peaks ($l=\text{integer}$) under Condition 1:
\begin{align*}
F_{\mathbf{G}=M\text{ peaks}}^j =& F_{l=\text{integer}}^j \\
=& 2\cos\Bigl(2\pi\mathbf{G}\cdot\Delta \mathbf{R}_1^j\Bigr) \\
&+ 2e^{i2\pi h}\cos\Bigl(2\pi\mathbf{G}\cdot\Delta \mathbf{R}_2^j\Bigr)\\
\quad &+ 2e^{i2\pi k}\cos\Bigl(2\pi\mathbf{G}\cdot\Delta \mathbf{R}_3^j\Bigr)\\
&+ 2e^{i2\pi (h+k)}\cos\Bigl(2\pi\mathbf{G}\cdot\Delta \mathbf{R}_4^j\Bigr).
\tag{S5}
\label{eq:S5}
\end{align*}

In general, $F_{l=\text{integer}}^j$ in the above equation is not necessarily zero, namely, the $M$ peaks are not strictly extinct. To ensure that the $M$ peaks are all extinct, we need to impose an additional constraint: for a given atom type $j$, the quantities $|\mathbf{G}\cdot\Delta \mathbf{R}_s^j|$ are identical for all subunit cells $s$. As we require $M$ peaks of all orders to be extinct, this condition suggests that for a given $j$,
\begin{align*}
\Delta \mathbf{R}_1^j = \pm \Delta \mathbf{R}_2^j = \pm \Delta \mathbf{R}_3^j = \pm \Delta \mathbf{R}_4^j,
\tag{S6}
\label{eq:S6}
\end{align*}
and we call this \textbf{Condition 2} (see Table~\ref{tab:L_phase_conditions}). Under this condition, Eq.~\eqref{eq:S5} is reduced to
\begin{align*}
F_{\mathbf{G}=M\text{ point}}^j =& F_{l=\text{integer}}^j
= 2\cos\Bigl(2\pi\mathbf{G}\cdot\Delta \mathbf{R}_1^j\Bigr)\\
&\times \Bigl[1+e^{i2\pi h} 
+e^{i2\pi k}+e^{i2\pi(h+k)}\Bigr] = 0,
\tag{S7}
\label{eq:S7}
\end{align*}
where the last equality utilizes the fact that at least one of $h$ and $k$ needs to be a half-integer for a superlattice peak so that the term in the bracket always evaluates to zero. 

We are now ready to consider the $L$ peaks in the $\mathcal{L}$ phase, which need to follow the set of extinction rules in Table~\ref{tab:extinction_rules}. Under Condition~1, the structure factor at $L$ points is reduced to:
\begin{align*}
F_{\mathbf{G}=L\text{ point}}^j =& F_{l=\text{half-integer}}^j \\
=& 2i\sin\Bigl(2\pi\mathbf{G}\cdot\Delta \mathbf{R}_1^j\Bigr)\\
&+ 2ie^{i2\pi h}\sin\Bigl(2\pi\mathbf{G}\cdot\Delta \mathbf{R}_2^j\Bigr)\\
&+ 2ie^{i2\pi k}\sin\Bigl(2\pi\mathbf{G}\cdot\Delta \mathbf{R}_3^j\Bigr)\\
&+ 2ie^{i2\pi (h+k)}\sin\Bigl(2\pi\mathbf{G}\cdot\Delta \mathbf{R}_4^j\Bigr).
\tag{S8}
\label{eq:S8}
\end{align*}

Applying Condition 2 to Eq.~\eqref{eq:S8} does not reproduce the $L$ peak extinction rules. It is necessary to further constrain the displacements along $\mathbf{a}$, $\mathbf{b}$, and $-(\mathbf{a}+\mathbf{b})$ directions with an anti-phase relation between adjacent subunit cells in specific directions, giving rise to a narrower set of conditions dubbed \textbf{Conditions 2a--c} (see Table~\ref{tab:L_phase_conditions}). One can readily verify that applying Conditions~2a, 2b, and 2c to Eq.~\eqref{eq:S8} separately reproduces a subset of the $L$ peak extinction rules, as summarized in Table~\ref{tab:L_phase_conditions}.

To reproduce all the observed extinction rules of the $L$ peaks, one needs to assign one and only one of Conditions~2a--c to each atomic type $j$ in one domain. In other words, for a specific type of atom in a single-crystalline superlattice, its displacement pattern cannot be a linear superposition of displacements corresponding to all three CDW wave vectors, as suggested earlier by the triple-$q$ structure \cite{DiSalvo1976}. Instead, the displacements microscopically break the three-fold rotational symmetry, and the underlying displacements are effectively one-dimensional for the sublattice of each atomic type Te, Se1, and Se2.

In any single-crystalline domain, there are now several possibilities to assign Ti, Se1, and Se2 to Conditions~2a--c, as summarized in Tables~\ref{tab:AG1}--\ref{tab:AG3}.

\begin{table}[ht]
\centering
\caption{\textbf{Assignment Group 1.} Each atomic type obeys one condition.}
 
\begin{tabularx}{0.8\columnwidth}{c|*{6}{X}}   
 
\Xhline{0.25ex}
\textbf{Condition} & \multicolumn{6}{c}{\textbf{Assignment to atomic types}} \\
\hline
2a & Ti  & Ti  & Se1 & Se2 & Se1 & Se2 \\
2b & Se1 & Se2 & Ti  & Ti  & Se2 & Se1 \\
2c & Se2 & Se1 & Se2 & Se1 & Ti  & Ti  \\
\Xhline{0.25ex}
\end{tabularx}

\label{tab:AG1}
\end{table}

\begin{table}[ht]
\centering
\caption{\textbf{Assignment Group 2.} Two atomic types obey the same condition.}
\begin{tabular}{c|cccccc}
\Xhline{0.25ex}
\textbf{Condition} & \multicolumn{6}{c}{\textbf{Assignment to atomic types}} \\
\hline
2a & Ti,Se1 & Ti,Se1 & Ti,Se2 & Ti,Se2 & Se1,Se2 & Se1,Se2 \\
2b & Se2    & --     & Se1    & --     & Ti      & --      \\
2c & --     & Se2    & --     & Se1    & --      & Ti      \\
\Xhline{0.25ex}
\end{tabular}
\label{tab:AG2-1}
\end{table}

\begin{table}[ht]
\centering
\caption{\textbf{Assignment Group 2.} (continued)}
\begin{tabular}{c|cccccc}
\Xhline{0.25ex}
\textbf{Condition} & \multicolumn{6}{c}{\textbf{Assignment to atomic types}} \\
\hline
2a & Se2   & --     & Se1    & --     & Ti      & --      \\
2b & Ti,Se1 & Ti,Se1 & Ti,Se2 & Ti,Se2 & Se1,Se2 & Se1,Se2 \\
2c & --     & Se2    & --     & Se1    & --      & Ti      \\
\Xhline{0.25ex}
\end{tabular}
\label{tab:AG2-2}
\end{table}

\begin{table}[H]
\centering
\caption{\textbf{Assignment Group 2.} (continued)}
\begin{tabular}{c|cccccc}
\Xhline{0.25ex}
\textbf{Condition} & \multicolumn{6}{c}{\textbf{Assignment to atomic types}} \\
\hline
2a & Se2   & --     & Se1    & --     & Ti      & --      \\
2b & --    & Se2    & --     & Se1    & --      & Ti      \\
2c & Ti,Se1 & Ti,Se1 & Ti,Se2 & Ti,Se2 & Se1,Se2 & Se1,Se2 \\
\Xhline{0.25ex}
\end{tabular}
\label{tab:AG2-3}
\end{table}

\begin{table}[H]
\centering
\caption{\textbf{Assignment Group 3.} All atomic types obey the same condition}
\begin{tabular}{c|ccc}
\Xhline{0.25ex}
\textbf{Condition} & \multicolumn{3}{c}{\textbf{Assignment to atomic types}} \\
\hline
2a & Ti,Se1,Se2 & -- & -- \\
2b & -- & Ti,Se1,Se2 & -- \\
2c & -- & -- & Ti,Se1,Se2 \\
\Xhline{0.25ex}
\end{tabular}
\label{tab:AG3}
\end{table}

Reconciling with the observations of the triple-$q$ structure in previous diffraction experiments, we interpret that it is likely all three wave vectors are equally present over macroscopic regions on the order of X-ray or neutron beam spots (tens of micrometers to millimeter scale), where the sample hosts a quasi-equal sampling of many coexisting domains. Our proposal offers an alternative explanation for the observation of all three wave vectors without invoking the microscopic superposition of displacement vectors on every atom in a conventional triple-$q$ structure. This explanation also accounts for the apparent three-fold symmetric diffraction pattern observed along the $[001]$ zone axis.

\section{S\lowercase{TRUCTURAL ANALYSIS OF THE }$\mathcal{M}$ \lowercase{PHASE}}

\begin{figure}[h]
    \centering
    \includegraphics[width=\columnwidth]{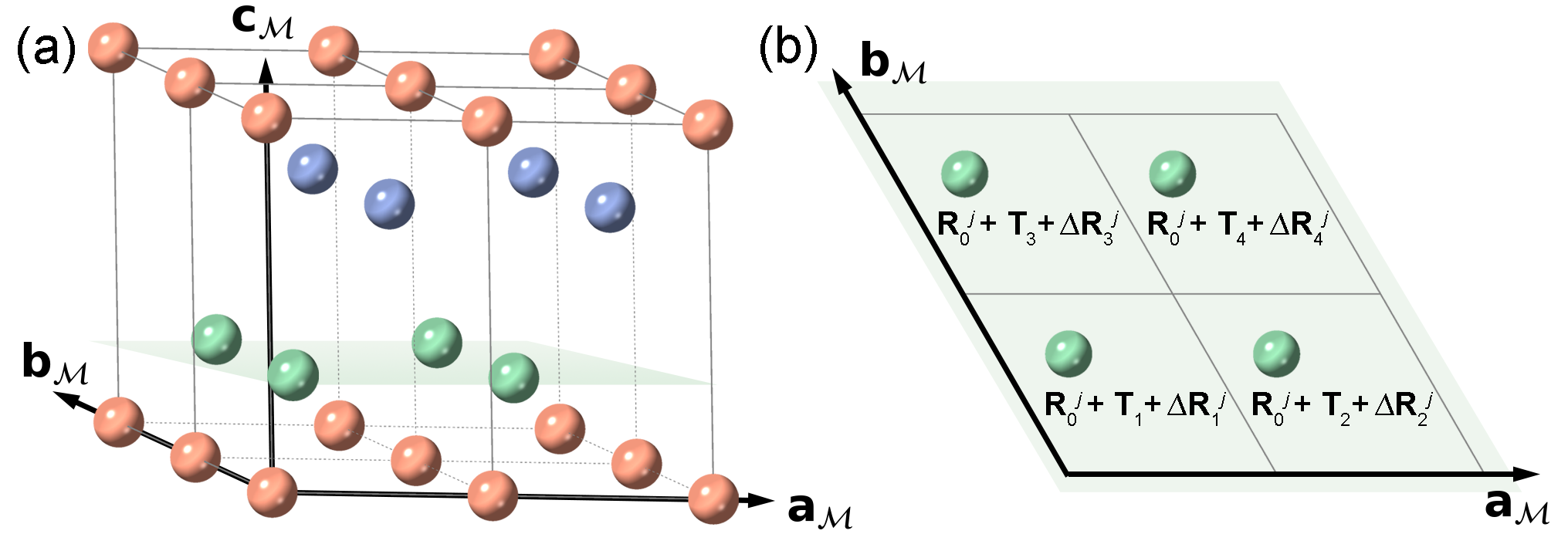}
    \caption{\justifying \textbf{Structure of the $2 \times 2 \times 1$ $\mathcal{M}$ phase superlattice.} (a) 3D schematic of the superlattice structure. (b) Top view of the supercell with Se1 shown.  See text and Fig.~\ref{fig:L_supercell} caption for definitions of $\mathbf{R}_0^j$, $\mathbf{T}_s$, $\Delta\mathbf{R}_s^j$ and $\mathbf{R}_s^j$. Here, $s = 1,2,3,4$.}
    \label{fig:M_supercell}
\end{figure}

\begin{figure}[h]
    \centering
    \includegraphics[width=\columnwidth]{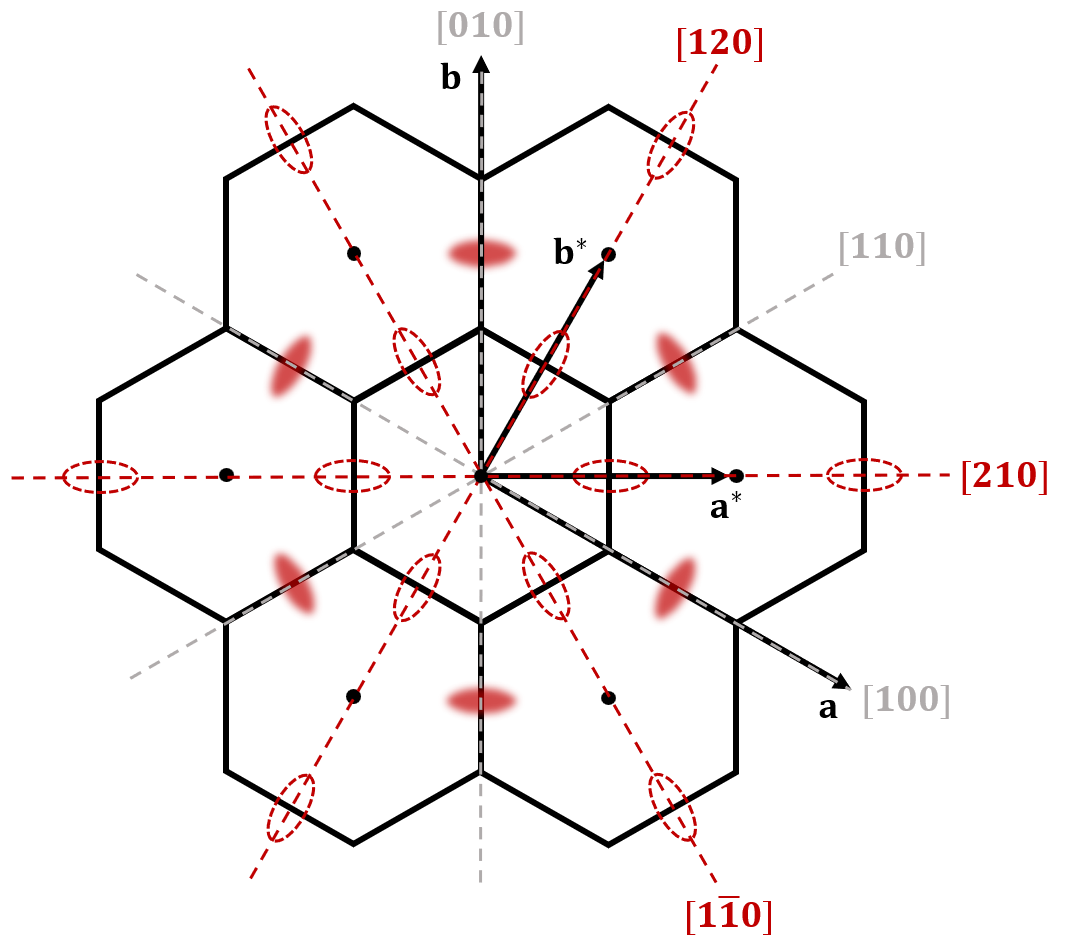}
    \caption{\justifying\textbf{$M$ peak extinction rules in the $\mathcal{M}$ phase.} Red dash lines represent zone axes along which $M$ peaks are visible in diffraction patterns, while gray dash lines represent those along which the $M$ peaks are extinct. The solid red ovals mark the visible $M$ peak, while the dashed red ovals mark the $M$ points where the peak is extinct.}
    \label{fig:M_peak_extinction_rules}
\end{figure}

In this section, we analyze the CDW atomic displacements associated with the $\mathcal{M}$ phase. For the $\mathcal{M}$ phase, $M$ peaks are not extinct, so we can choose a $2\times2\times1$ supercell to incorporate the structure, as illustrated in Fig.~\ref{fig:M_supercell}. 
Table~\ref{tab:M_peak_extinction_rules} and Fig.~\ref{fig:M_peak_extinction_rules} summarize the experimental observations of $M$ peaks in the diffraction patterns along various zone axes and the associated extinction rules. We will work out the conditions for the atomic displacements that satisfy these extinction rules.

\begin{table}[ht]
    \centering
    \caption{\justifying\textbf{Present and extinct $M$ peaks in diffraction patterns along various zone axes.} Red text indicates zone axes along which $M$ peaks are present; gray text indicates zone axes along which $M$ peaks are extinct.}
    \begin{tabular}{c|c}
    \Xhline{0.25ex}
    \textbf{Zone axis} & \textbf{Present $M$ peaks} \\
    \hline
    \textcolor{BrickRed}{$[1\overline{1}0]\quad$} & \textcolor{BrickRed}{$(h, h, l)$} \\
    \textcolor{BrickRed}{$[120]\quad$} & $\textcolor{BrickRed}{(2h, \overline{h}, l)}$ \\
    \textcolor{BrickRed}{$[210]\quad$} & $\textcolor{BrickRed}{(h, \overline{2h}, l)}$ \\
    \Xhline{0.25ex}
    \textbf{ } & \textbf{Extinct $M$ peaks} \\
    \hline
    \textcolor{gray}{$[010]\quad$} & \textcolor{gray}{$(h, 0, l)$} \\
    \textcolor{gray}{$[100]\quad$} & \textcolor{gray}{$(0, k, l)$} \\
    \textcolor{gray}{$[110]\quad$} & \textcolor{gray}{$(h, \overline{h}, l)$} \\
    \Xhline{0.25ex}
    \end{tabular}
    \label{tab:M_peak_extinction_rules}
\end{table}

Similar to $\mathcal{L}$, the superlattice structure factor of $\mathcal{M}$ phase is given by:
\begin{align*}
F_{\mathbf{G}} 
=& \sum_{j=1}^{3} \sum_{s=1}^{4} f_j  e^{i2\pi\mathbf{G}\cdot \bigl(\mathbf{R}_0^j + \mathbf{T}_s + \Delta \mathbf{R}_s^j\bigr)} \\
=& \sum_{j=1}^{3} f_j  F_{\mathbf{G}}^j  e^{i2\pi\mathbf{G}\cdot \mathbf{R}_0^j},
\tag{S9}
\label{eq:S9}
\end{align*}
where $\mathbf{G} = 2h\mathbf{a}_{\mathcal{M}}^* + 2k\mathbf{b}_{\mathcal{M}}^* + l\mathbf{c}_{\mathcal{M}}^*$ for the $\mathcal{M}$ reciprocal superlattice vectors 
$\mathbf{a}_{\mathcal{M}}^*, \mathbf{b}_{\mathcal{M}}^*, \mathbf{c}_{\mathcal{M}}^*$. $(2h, 2k, l)$ are superlattice peaks in the lattice vector basis of the $2\times 2\times 1$ supercell. Here, at least one of $h$ or $k$ is a half-integer, and $l$ is an integer.

Let us consider both the real and imaginary components of the structure factor $F_{\mathbf{G}}^j$. 
Its real component is

\begin{align*}
\mathrm{Re}\left[F_{\mathbf{G}}^j\right]
=& \cos\bigl(2\pi\mathbf{G}\cdot \Delta\mathbf{R}_1^j\bigr) \\
&+ e^{i2\pi h}\cos\bigl(2\pi\mathbf{G}\cdot \Delta\mathbf{R}_2^j\bigr) \\
&+ e^{i2\pi k}\cos\bigl(2\pi\mathbf{G}\cdot \Delta\mathbf{R}_3^j\bigr) \\
&+ e^{i2\pi (h + k)}\cos\bigl(2\pi\mathbf{G}\cdot \Delta\mathbf{R}_4^j\bigr).
\tag{S10}
\label{eq:S10}
\end{align*}
We note that Eq.~\eqref{eq:S10} is the same as Eq.~\eqref{eq:S5} up to a global factor of 2, so Condition~2 from Table~\ref{tab:M_phase_conditions} enforces $\mathrm{Re}\left[F_{\mathbf{G}}^j\right] = 0$ for all $M$ peaks.

For the imaginary component of $F_{\mathbf{G}}^j$:
\begin{align*}
\mathrm{Im}\left[F_{\mathbf{G}}^j\right]
=& \sin\bigl(2\pi \mathbf{G} \cdot \Delta \mathbf{R}_1^j\bigr) \\
&+ e^{i2\pi h}\sin\bigl(2\pi \mathbf{G} \cdot \Delta \mathbf{R}_2^j\bigr) \\
&+ e^{i2\pi k}\sin\bigl(2\pi \mathbf{G} \cdot \Delta \mathbf{R}_3^j\bigr) \\
&+ e^{i2\pi (h + k)}\sin\bigl(2\pi \mathbf{G} \cdot \Delta \mathbf{R}_4^j\bigr),
\tag{S11}
\label{eq:S11}
\end{align*}
which is the same as Eq.~\eqref{eq:S8} up to a global factor of $2i$. Hence, the same conclusion about the $\mathcal{L}$ phase analysis follows. Namely, applying Conditions~2a--c to Eq.~\eqref{eq:S11} for Ti, Se1, and Se2 respectively yields the correct $M$ peak extinction rules, as summarized in Table~\ref{tab:M_phase_conditions}, and the underlying displacement for the sublattice of each atomic type is one-dimensional.

Due to the anisotropic shape of the peaks at the $M$ point, it is further needed to enforce the following:
\begin{quote}
\textit{For any given atomic type $j$, the correlation length of the distortion parallel to the underlying wave vector of Condition~2a/2b/2c is shorter than the correlation length perpendicular to the wave vector.}
\end{quote}
This condition reproduces the elongation of the peaks at the $M$ point in the $\mathrm{\Gamma}$--$M$ direction. As discussed in the main text, this anisotropy follows naturally from the one-dimensional displacement pattern for each tomic type.

Similar to the $\mathcal{L}$ phase, there is also likely a quasi-equal distribution of domains resulting from different ways of assigning Conditions~2a, 2b, and 2c to the atomic types Te, Se1, and Se2.

\section{S\lowercase{ymmetry analysis of distinct domains}}

In the main text, we described how the three CDW wave vectors of 1$T$-TiSe$_2$ could be assigned to the three atomic types combinatorially, giving rise to 27 primary domains. Here, we calculate the space groups of each primary domain.

The well-established high-temperature structure of 1$T$-TiSe$_2$ is imported from a refined structure \cite{Kitou2019}, and then expanded into a $2\times2\times2$ or $2\times2\times1$ supercell to accommodate the $\mathcal{L}$ and $\mathcal{M}$ phases, respectively. Displacements with amplitudes assumed from Di Salvo's structure \cite{DiSalvo1976} were applied to different atomic types (Ti, Se1, Se2) according to each permutation of the wave vectors, denoted as tuples $(\mathbf{q}_i$~$\mathbf{q}_j$~$\mathbf{q}_k)$ for $i, j, k \in \{1, 2, 3\}$. Note that the one-dimensional displacements technically correspond to the in-plane component of the wave vectors $\mathbf{q}^{\parallel}$, but the $\parallel$ superscript is suppressed here to avoid clutter. The space group of each displaced structure was evaluated using the \texttt{spglib} package \cite{spglib}. The results are listed in Table~\ref{tab:L_phase_SGs} for the $\mathcal{L}$ phase and Table~\ref{tab:M_phase_SGs} for the $\mathcal{M}$ phase. ``Multiplicity" indicates the number of wave vector permutations that satisfy the general form listed in the ``wave vectors" column. These results were corroborated by calculating the space group of the displaced structures on Bilbao Crystallographic Server \cite{Bilbao}. 

\begin{table}[t!]
\caption{\justifying\textbf{Space groups of 27 $\mathcal{L}$ phase domains.} ($\mathbf{q}_i$~$\mathbf{q}_j$~$\mathbf{q}_k$) for $i, j, k \in \{1, 2, 3\}$ denotes the wave vectors corresponding to the displacements of the three atomic types of the parent lattice, (Ti, Se1, Se2). In the following, $\parallel$ superscript is suppressed and $i\neq j\neq k$.}
\centering
\label{tab:L_phase_SGs}
\begin{tabular}{c|c|c|c}
\Xhline{0.25ex}
~wave vectors~ & ~Multiplicity~ & ~Space group~ & ~Chirality~ \\
\hline
  ($\mathbf{q}_i$  $\mathbf{q}_i$  $\mathbf{q}_i$) &  3  &    $C2/c$ (No.\,15) & Not Chiral \\
  ($\mathbf{q}_i$  $\mathbf{q}_j$  $\mathbf{q}_j$) &  6  &    $P\overline{1}$ (No.\,2) & Not Chiral \\
  ($\mathbf{q}_i$  $\mathbf{q}_i$  $\mathbf{q}_j$) &  6  &    $P1$ (No.\,1) &     Chiral \\
  ($\mathbf{q}_i$  $\mathbf{q}_j$  $\mathbf{q}_i$) &  6  &    $P1$ (No.\,1) &     Chiral \\
  ($\mathbf{q}_i$  $\mathbf{q}_j$  $\mathbf{q}_k$) &  6  &    $C2$ (No.\,5) &     Chiral \\
\Xhline{0.25ex}
\end{tabular}
\end{table}

\begin{table}[t!]
\caption{\justifying\textbf{Space Groups of 27 $\mathcal{M}$ phase domains.} The notation here follows the convention in Table~\ref{tab:L_phase_SGs}.}
\centering
\label{tab:M_phase_SGs}
\begin{tabular}{c|c|c|c}
\Xhline{0.25ex}
~wave vectors~ & ~Multiplicity~ & ~Space group~ & ~Chirality~ \\
\hline
  ($\mathbf{q}_i$  $\mathbf{q}_i$  $\mathbf{q}_i$) &  3  &    $Pc$ (No.\,7) & Not Chiral \\
  ($\mathbf{q}_i$  $\mathbf{q}_j$  $\mathbf{q}_j$) &  6  &    $P\overline{1}$ (No.\,2) & Not Chiral \\
  ($\mathbf{q}_i$  $\mathbf{q}_i$  $\mathbf{q}_j$) &  6  &    $P1$ (No.\,1) &     Chiral \\
  ($\mathbf{q}_i$  $\mathbf{q}_j$  $\mathbf{q}_i$) &  6  &    $P1$ (No.\,1) &     Chiral \\
  ($\mathbf{q}_i$  $\mathbf{q}_j$  $\mathbf{q}_k$) &  6  &    $C2$ (No.\,5) &     Chiral \\
\Xhline{0.25ex}
\end{tabular}
\end{table}

Importantly, for both $\mathcal{L}$ and $\mathcal{M}$ phases, 18 out of the 27 primary domains belong to either the $P1$ or $C2$ groups, which are chiral for breaking all of inversion, mirror, and rotoinversion symmetries \cite{Wagniere2007}. The space group $P1$ has been proposed for the CDW superlattice of 1$T$-TiSe$_2$ from a combination of X-ray structural refinement and Raman scattering \cite{Kim2024}, while the space group $C2$ has been proposed in scanning tunneling microscopy \cite{Kim2024b} and density functional perturbation theory \cite{Subedi2022} studies. Although details of the structures proposed in these studies are different from our proposal, our results offer a possible framework to explain why many different space groups arose in literature for the CDW superlattice of 1$T$-TiSe$_2$. Namely, the domain composition of each sample could differ from experiment to experiment due to different local strain environment, cooling history, and Se vacancy levels, resulting in different symmetry properties.

\begin{sidewaystable*}[htbp]
\vspace{3in}
\caption{\justifying\textbf{Conditions on atomic displacements in the $\mathcal{L}$ phase}. The conditions are shown with schematics of atomic displacements and the resulting extinction rules when applied to Eq.~\eqref{eq:S4}. All extinction rules refer to $L$ peaks, except in row~2. In the ``Atomic displacement schematics" column, only one atomic type (Se2) is displayed for Conditions~2,~2a--c for visual clarity; for Condition~3, only one out of 216 maximally possible displacement patterns is shown. The numbers labeled on the atoms correspond to the subunit cell index $s$, where $s=1,2,3,4$ and $s=5,6,7,8$ indicate the bottom- and top-layer subunit cells, respectively. The blue dots in ``Superlattice peak extinction rules schematics" represent non-vanishing $L$ peaks.}
\centering
\begin{adjustbox}{max width=\textwidth}  
\fontsize{9}{11}\selectfont  
\begin{tabular}{c|c|c|c|c|c}
    \Xhline{0.25ex}
    Conditions & Expressions & Atomic displacement schematics & Present superlattice Peaks & Extinct superlattice Peaks & Superlattice peaks extinction rules schematics \\
     
    \Xhline{0.25ex}
    1 & \( \Delta \mathbf{R}_{s+4}^j = -\Delta \mathbf{R}_s^j \) &
    \begin{minipage}{8.5cm}
      \centering
            \includegraphics[width=7.9cm,height=3.3cm,trim=0.0cm 0.0cm 0.0cm 0.0cm, clip]{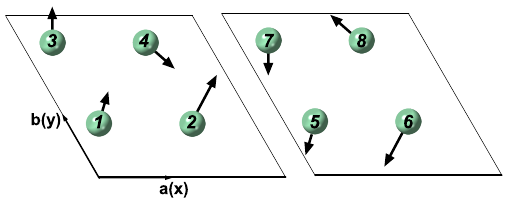}
    \end{minipage} &  N/A & N/A & N/A  \\
    \hline
    
    2 & \( \Delta \mathbf{R}_1^j = \pm \Delta \mathbf{R}_2^j = \pm \Delta \mathbf{R}_3^j = \pm \Delta \mathbf{R}_4^j \)&
       \begin{minipage}{8.5cm}
      \centering
            \includegraphics[width=7.9cm,height=3.3cm,trim=0.1cm 0.1cm 0.1cm 0.1cm, clip]{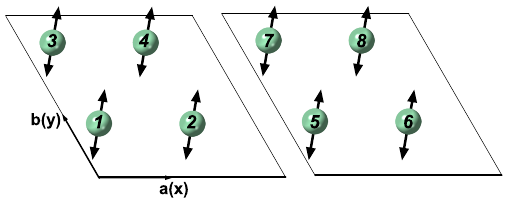}
    \end{minipage} & 
    
      \(
      \begin{array}{c}
      \text{$L$ peaks:} \\
       {(h, \overline{2h}, l)}, {(2h, \overline{h}, l)}, {(h, h, l)} \\
        {(h, \overline{h}, l)}, {(h, 0, l)}, {(0, h, l)} \\
      \end{array}
      \) & 
      \(
      \begin{array}{c}
      \text{$M$ peaks:} \\
       {(h, \overline{2h}, l)}, {(2h, \overline{h}, l)}, {(h, h, l)} \\
        {(h, \overline{h}, l)}, {(h, 0, l)}, {(0, h, l)} \\
        \text{(All $M$ peaks are extinct}\\
        \text{after this row)}
      \end{array}
      \) & 
      \begin{minipage}{8.5cm}
      \centering
            \includegraphics[width=5.2cm,height=3.3cm,trim=0.0cm 0.0cm 0.0cm 0.0cm, clip]{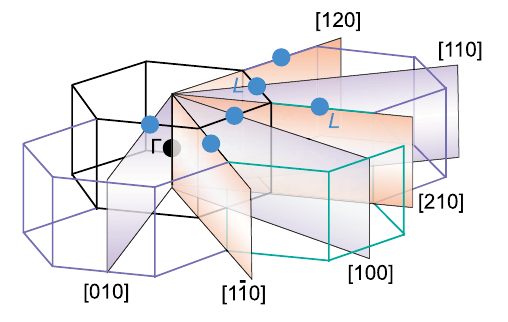}
    \end{minipage} \\
    \hline
    
    2a & \(
        \begin{array}{c}
          \Delta R_{3,y}^j = \Delta R_{1,y}^j \\ 
          \Delta R_{4,y}^j = \Delta R_{2,y}^j = -\Delta R_{1,y}^j\\
          \Delta R_{s,z}^j = \Delta R_{s,x}^j = 0\\   
        \end{array}
    \) 
      & 
      \begin{minipage}{8.5cm}
      \centering
            \includegraphics[width=7.9cm,height=3.3cm,trim=0.0cm 0.0cm 0.0cm 0.0cm, clip]{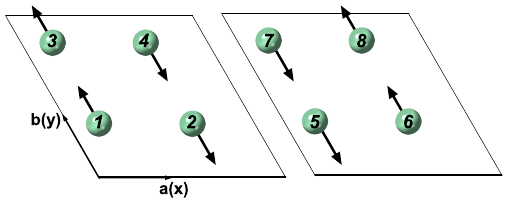}
    \end{minipage} & 
    \(
      \begin{array}{c}
       {(h, \overline{2h}, l)} \\
      \end{array}
      \)&  
      \(
      \begin{array}{c}
        {(2h, \overline{h}, l)}, {(h, h, l)} \\
        {(h, \overline{h}, l)}, {(h, 0, l)}, {(0, h, l)} \\
      \end{array}
      \) & 
       \begin{minipage}{8.5cm}
      \centering
            \includegraphics[width=5.2cm,height=3.3cm,trim=0.0cm 0.0cm 0.0cm 0.0cm, clip]{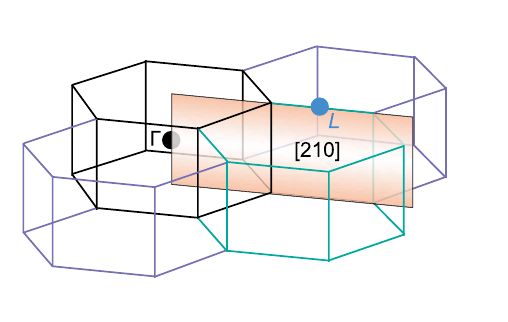}
    \end{minipage} \\
    \hline
   
    2b & \(
        \begin{array}{c}
          \Delta R_{2,x}^j = \Delta R_{1,x}^j \\ 
          \Delta R_{4,x}^j = \Delta R_{3,x}^j = -\Delta R_{1,x}^j\\
          \Delta R_{s,z}^j = \Delta R_{s,y}^j = 0\\   
        \end{array}
    \) 
      & 
       \begin{minipage}{8.5cm}
      \centering
            \includegraphics[width=7.9cm,height=3.3cm,trim=0.0cm 0.0cm 0.0cm 0.0cm, clip]{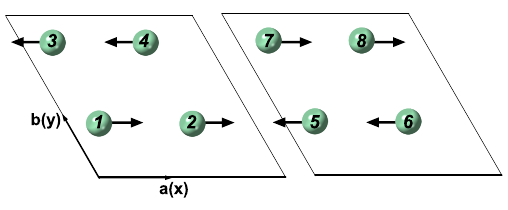}
    \end{minipage} & 
    \(
      \begin{array}{c}
      
       {(2h, \overline{h}, l)} \\
      \end{array}
      \)& 
      \(
      \begin{array}{c}
      
        {(h, \overline{2h}, l)}, {(h, h, l)} \\
        {(h, \overline{h}, l)}, {(h, 0, l)}, {(0, h, l)} \\
      \end{array}
      \) &   
      \begin{minipage}{8.5cm}
      \centering
            \includegraphics[width=5.2cm,height=3.3cm,trim=0.0cm 0.0cm 0.0cm 0.0cm, clip]{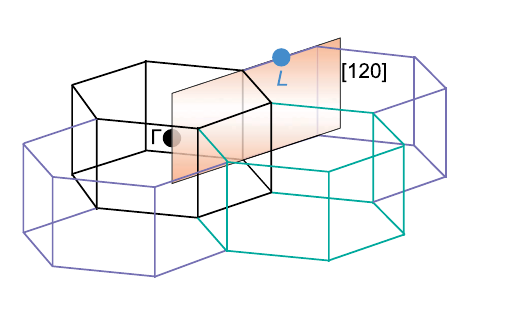}
    \end{minipage} \\
    \hline

    2c & \(
        \begin{array}{c}
          \Delta R_{4,y}^j = \Delta R_{4,x}^j = \Delta R_{1,y}^j = \Delta R_{1,x}^j \\ 
          \Delta R_{2,y}^j = \Delta R_{2,x}^j = \Delta R_{3,y}^j = \Delta R_{3,x}^j \\
           = -\Delta R_{1,x}^j\\
          \Delta R_{s,z}^j = 0\\   
        \end{array}
    \)  
      & 
       \begin{minipage}{8.5cm}
      \centering
            \includegraphics[width=7.9cm,height=3.3cm,trim=0.0cm -0.1cm 0.0cm 0.0cm, clip]{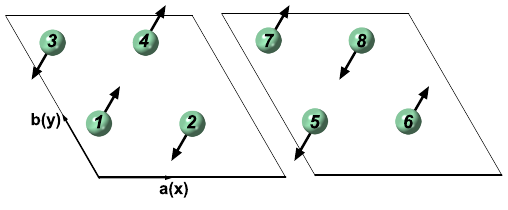}
    \end{minipage} & 
    \(
      \begin{array}{c}
      
       {(h, h, l)} \\
      \end{array}
      \)&  
      \(
      \begin{array}{c}
      
        {(h, \overline{2h}, l)}, {(2h, \overline{h}, l)} \\
        {(h, \overline{h}, l)}, {(h, 0, l)}, {(0, h, l)} \\
      \end{array}
      \) & 
        \begin{minipage}{8.5cm}
      \centering
            \includegraphics[width=5.2cm,height=3.3cm,trim=0.0cm 0.0cm 0.0cm 0.0cm, clip]{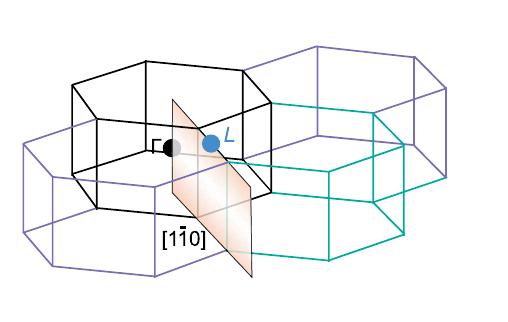}
    \end{minipage} \\
    \hline
    
    3 &  \makecell{2a, 2b, 2c assigned to \\  Ti, Se1, Se2} & 
      \begin{minipage}{8.5cm}
      \centering
            \includegraphics[width=8.3cm,height=3.3cm]{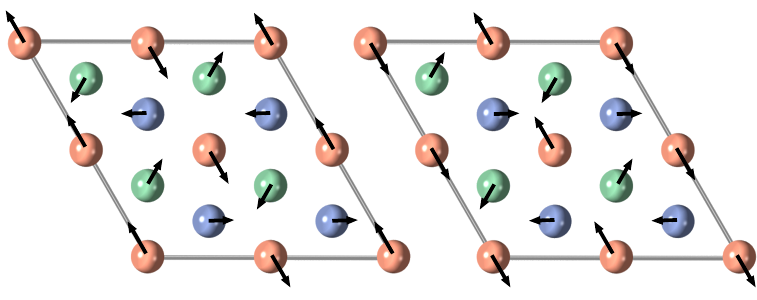}
    \end{minipage} & 
     \(
      \begin{array}{c}
      
       {(h, \overline{2h}, l)}, {(2h, \overline{h}, l)}, {(h, h, l)} \\
      \end{array}
      \) & 
      \(
      \begin{array}{c}
      
        {(h, \overline{h}, l)}, {(h, 0, l)}, {(0, h, l)} \\
         
      \end{array}
      \) & 
        \begin{minipage}{8.5cm}
      \centering
            \includegraphics[width=5.2cm,height=3.3cm,trim=0.0cm 0.0cm 0.0cm 0.0cm, clip]{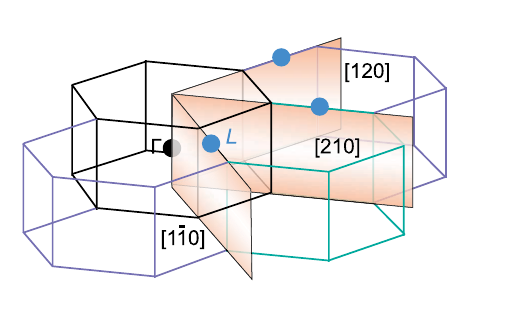}
    \end{minipage} \\
    \Xhline{0.25ex}
    
\end{tabular}
\end{adjustbox}

\label{tab:L_phase_conditions}
\end{sidewaystable*}

\begin{sidewaystable*}[htbp]
\vspace{3in}
\caption{\justifying\textbf{Conditions on atomic displacements in the $\mathcal{M}$ phase.} The conditions are shown with the resulting extinction rules for the $M$ peaks when they are applied to Eq.~\eqref{eq:S9}. All $L$ peaks are extinct because we have assumed no unit cell doubling along $c^*$ in the $\mathcal{M}$ phase. The red dots represent non-vanishing $M$ peaks. See Table~\ref{tab:L_phase_conditions} caption for a detailed description.} 
\centering
\begin{adjustbox}{max width=\textwidth}  
\fontsize{9}{11}\selectfont  
\begin{tabular}{c|c|c|c|c|c}
    \Xhline{0.25ex}
     Conditions & Expressions & Atomic displacement schematics & Allowed superlattice Peaks & Extinct superlattice Peaks & Superlattice peaks extinction rules schematics \\
    \Xhline{0.25ex}
    1 & N/A & N/A & N/A & N/A & N/A \\
    \hline
    2 & \( \Delta \mathbf{R}_1^j = \pm \Delta \mathbf{R}_2^j = \pm \Delta \mathbf{R}_3^j = \pm \Delta \mathbf{R}_4^j \)&
    \begin{minipage}[c]{8.5cm}
      \centering
            \includegraphics[width=4.5cm,height=3.3cm,trim=0.0cm 0.0cm 0.0cm 0.0cm, clip]{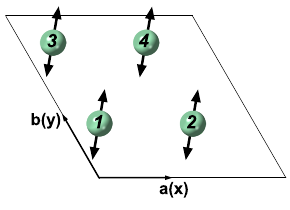}
    \end{minipage} & 
      \(
      \begin{array}{c}
       {(h, \overline{2h}, l)}, {(2h, \overline{h}, l)}, {(h, h, l)} \\
        {(h, \overline{h}, l)}, {(h, 0, l)}, {(0, h, l)} \\
      \end{array}
      \) & N/A & 
      \begin{minipage}{8.5cm}
      \centering
            \includegraphics[width=5.2cm,height=3.3cm,trim=0.0cm 0.0cm 0.0cm 0.0cm, clip]{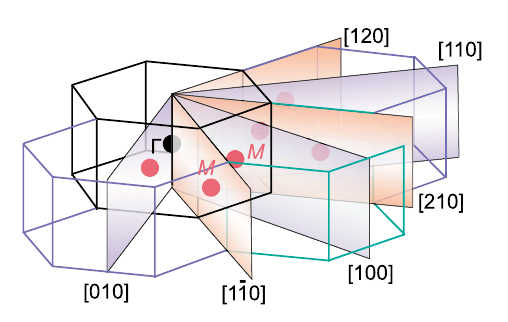}
    \end{minipage} \\
    \hline
    
    2a & \(
        \begin{array}{c}
          \Delta R_{3,y}^j = \Delta R_{1,y}^j \\ 
          \Delta R_{4,y}^j = \Delta R_{2,y}^j = -\Delta R_{1,y}^j\\
          \Delta R_{s,z}^j = \Delta R_{s,x}^j = 0\\   
        \end{array}
    \) 
      & 
      \begin{minipage}[c]{8.5cm}
      \centering
            \includegraphics[width=4.5cm,height=3.3cm,trim=0.0cm 0.0cm 0.0cm 0.0cm, clip]{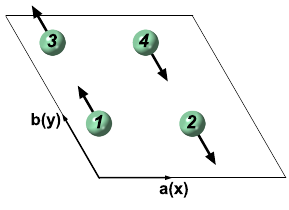}
    \end{minipage} & 
      
    \(
      \begin{array}{c}
      
       {(h, \overline{2h}, l)} \\
      \end{array}
      \)&  
      \(
      \begin{array}{c}
        {(2h, \overline{h}, l)}, {(h, h, l)} \\
        {(h, \overline{h}, l)}, {(h, 0, l)}, {(0, h, l)} \\
      \end{array}
      \) & 
       \begin{minipage}{8.5cm}
      \centering
            \includegraphics[width=5.2cm,height=3.3cm,trim=0.0cm 0.0cm 0.0cm 0.0cm, clip]{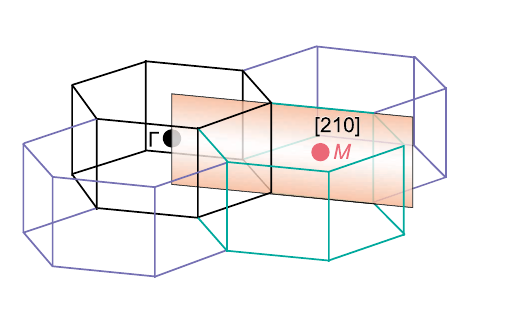}
    \end{minipage} \\
    \hline
   
    2b & \(
        \begin{array}{c}
          \Delta R_{2,x}^j = \Delta R_{1,x}^j \\ 
          \Delta R_{4,x}^j = \Delta R_{3,x}^j = -\Delta R_{1x}^j\\
          \Delta R_{s,z}^j = \Delta R_{s,y}^j = 0\\   
        \end{array}
    \) 
      & 
      \begin{minipage}[c]{8.5cm}
      \centering
            \includegraphics[width=4.5cm,height=3.3cm,trim=0.0cm 0.0cm 0.0cm 0.0cm, clip]{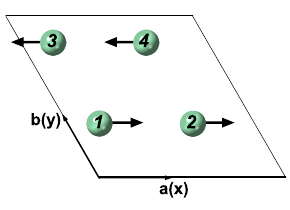}
    \end{minipage} & 
    
    \(
      \begin{array}{c}
      
       {(2h, \overline{h}, l)} \\
      \end{array}
      \)& 
      \(
      \begin{array}{c}
     
        {(h, \overline{2h}, l)}, {(h, h, l)} \\
        {(h, \overline{h}, l)}, {(h, 0, l)}, {(0, h, l)} \\
      \end{array}
      \) &   
      \begin{minipage}{8.5cm}
      \centering
            \includegraphics[width=5.2cm,height=3.3cm,trim=0.0cm 0.0cm 0.0cm 0.0cm, clip]{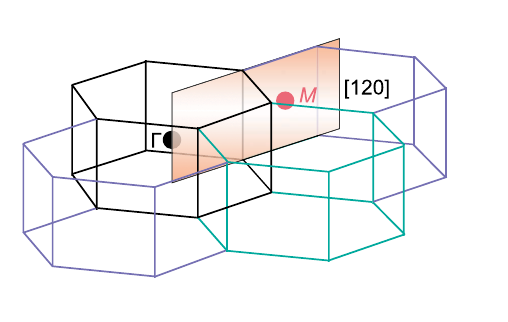}
    \end{minipage} \\
    \hline

    2c & \(
        \begin{array}{c}
          \Delta R_{4,y}^j = \Delta R_{4,x}^j = \Delta R_{1,y}^j = \Delta R_{1,x}^j \\ 
          \Delta R_{2,y}^j = \Delta R_{2,x}^j = \Delta R_{3,y}^j = \Delta R_{3,x}^j \\
           = -\Delta R_{1,x}^j\\
          \Delta R_{s,z}^j = 0\\   
        \end{array}
    \)  
      & 
      \begin{minipage}[c]{8.5cm}
      \centering
            \includegraphics[width=4.5cm,height=3.3cm,trim=0.0cm 0.0cm 0.0cm 0.0cm, clip]{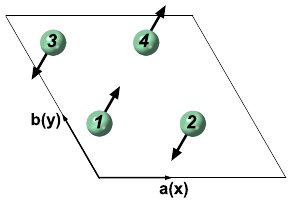}
    \end{minipage} & 
       
    \(
      \begin{array}{c}
     
       {(h, h, l)} \\
      \end{array}
      \)&  
      \(
      \begin{array}{c}
        {(h, \overline{2h}, l)}, {(2h, \overline{h}, l)} \\
        {(h, \overline{h}, l)}, {(h, 0, l)}, {(0, h, l)} \\
      \end{array}
      \) & 
        \begin{minipage}{7.5cm}
      \centering
            \includegraphics[width=5.2cm,height=3.3cm,trim=0.0cm 0.0cm 0.0cm 0.0cm, clip]{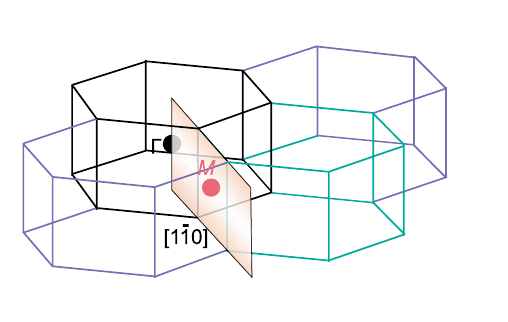}
    \end{minipage} \\
    \hline
    
    3 &  \makecell{2a, 2b, 2c assigned to \\  Ti, Se1, Se2} &  
      \begin{minipage}[c]{8.5cm}
      \centering
            \includegraphics[width=4.5cm,height=3.3cm,trim=0.0cm 0.0cm 0.0cm 0.0cm, clip]{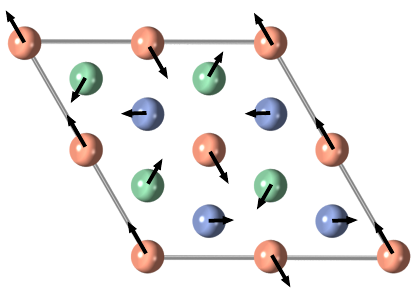}
    \end{minipage} & 
     \(
      \begin{array}{c}
       {(h, \overline{2h}, l)}, {(2h, \overline{h}, l)}, {(h, h, l)} \\
      \end{array}
      \) & 
      
      \(
      \begin{array}{c}
        {(h, \overline{h}, l)}, {(h, 0, l)}, {(0, h, l)} \\
        
      \end{array}
      \) & 
        \begin{minipage}{8.5cm}
      \centering
            \includegraphics[width=5.2cm,height=3.3cm,trim=0.0cm 0.0cm 0.0cm 0.0cm, clip]{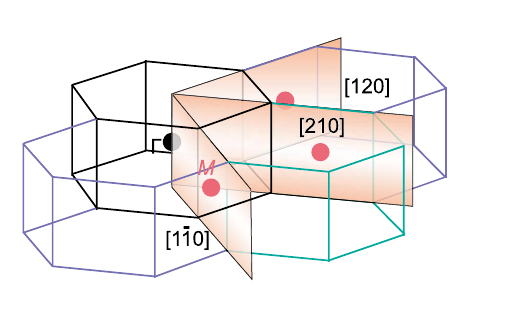}
    \end{minipage} \\
    \Xhline{0.25ex}
    
\end{tabular}
\end{adjustbox}

\label{tab:M_phase_conditions}
\end{sidewaystable*}

\makeatletter\@input{xx.tex}\makeatother